\documentclass[twocolumn,showpacs,preprintnumbers,amsmath,amssymb, nofootinbib]{revtex4}

\usepackage{pdflscape}
\usepackage{color}
\usepackage{dcolumn}
\usepackage{bm}
\usepackage{epsfig}%
\usepackage{amsmath}
\usepackage[colorlinks]{hyperref}
\usepackage{ulem}
\newcommand{\tb}{\textcolor{black}}
\newcommand{\tr}{\textcolor{black}}
\newcommand{\tg}{\textcolor{black}}

\begin{document}

\title{Meissner Effect: History of Development and Novel Aspects}



\author{Vladimir Kozhevnikov }
\affiliation{Tulsa Community College, Tulsa, Oklahoma 74119, USA}.  \\
\affiliation{KU Leuven, BE-3001 Leuven, Belgium}.  \\


\begin{abstract}
\noindent The discovery of the Meissner (Meissner–Ochsenfeld) effect in 1933 was an incontestable turning point in the history of superconductivity.  First, it demonstrated that superconductivity is an unknown before equilibrium state of matter, thus allowing to use the power of thermodynamics for its study. This provided a justification for the two-fluid model of Gorter and Casimir, a seminal thermodynamic theory founded on a postulate of zero entropy of the superconducting (S) component of conduction electrons. Second, the Meissner effect demonstrated that, apart from zero electric resistivity,  the S phase is also characterized by zero magnetic induction. The latter property is used as a basic postulate in the theory of F. and H. London, which underlies the  understanding of electromagnetic properties of superconductors. Here the experimental and theoretical aspects of the Meissner effect are reviewed.  The reader will see that, in spite of almost nine decades age, the London theory still contains questions, the answers to which  can lead to a revision of the standard picture of the  Meissner state (MS) and, if so, of other equilibrium superconducting states. An attempt is made to take a fresh look at  electrodynamics of the MS and try to work out with the issues associated with \tg{the description of this most} important state of all superconductors. It is shown that the concept of Cooper's pairing along with the Bohr–Sommerfeld quantization condition allows one to construct a semi-classical theoretical model consistently addressing properties of the MS and beyond, including non-equilibrium properties of superconductors caused by the total current. As follows from the model, the three “big zeros” of superconductivity (zero resistance, zero induction and zero entropy) have equal weight and grow from a single root: quantization of the angular momentum of paired electrons. The model predicts some \tg{novel} 
effects. If confirmed, they can help in studies of microscopic properties  of all superconductors. Preliminary experimental results suggesting the need to revise the  standard picture of the MS are presented. 
 
\end{abstract}\

\maketitle  

\hspace{15mm}Contents\vspace{1.2mm}

\hspace{6.5mm}\tb{Frequently used abbreviations}\vspace{0.8mm} 

I.\hspace{3mm} MEISSNER EFFECT\vspace{0.8mm}

\hspace{10mm}Meissner and Ochsenfeld\vspace{0.8mm}

\hspace{10mm}Rjabinin and Shubnikov\vspace{1mm}

II.\hspace{1.5mm} MEISSNER STATE DEFINITION\vspace{1mm}

III.\hspace{0.5mm} TWO-FLUID MODEL\vspace{1mm}

IV.\hspace{0.5mm} LONDON THEORY\vspace{1mm}

V.\hspace{2mm} MICRO-WHIRLS MODEL\vspace{0.8mm}

\hspace{9mm}{\small PROPERTIES OF THE MEISSNER STATE}\vspace{0.8mm}

\hspace{9mm}{\small FLUX QUANTIZATION}\vspace{0.8mm}

\hspace{9mm}{\small OTHER PROPERTIES}\vspace{0.8mm}

\hspace{11mm}Hall effect\vspace{0.8mm}

\hspace{11mm}Paramagnetism of the Abrikosov vortices\vspace{0.8mm}

\hspace{11mm}Surface tension\vspace{0.8mm}

\hspace{11mm}Type-I to type-II conversion\vspace{0.8mm}

\hspace{11mm}Total current\vspace{1mm}

VI.\hspace{1mm} EXPERIMENT\vspace{1mm}

\hspace{0mm} SUMMARY AND OUTLOOK\vspace{1mm}

\hspace{0mm} APPENDIX\vspace{1mm}

\hspace{0mm} ACKNOWLEDGMENTS\vspace{1mm}

\hspace{0mm} References

\hspace{4mm} \tb{Frequently used abbreviations}\vspace{1mm}

\tb{BCS - Bardeen-Cooper-Schrieffer (theory)}

\tb{GL \hspace{2mm}- Ginzburg-Landau (theory)}

\tb{MW \hspace{0mm}- micro-whirls (model)} 

\tb{MS \hspace{2mm}- Meissner state}

\tb{N}  \hspace{4mm}- normal (state, phase, \textit{etc.})

\tb{S} \hspace{5mm}- superconducting (state, phase, \textit{etc.})

\section{MEISSNER EFFECT}

The history of the Meissner effect takes its origin  from experiments of Keesom and coworkers of 1932 \cite{Keesom_1932}, in which it was revealed that the electron heat capacity in tin and thallium experiences a discontinuous jump near the critical temperature of the transition from the normal (N) to the superconducting (S) state $T_c$, a constant of the material in question. 

Previous two decades after the discovery of superconductivity in the laboratory of Kamerlingh Onnes \cite{Onnes_1911} superconductors were viewed as perfect (resistanceless) conductors. This implies that a dc electric current set in a closed superconducting circuit will be running persistently, i.e. without any kind of energy dissipation like it occurs with electrons bound in atoms. Such a viewpoint was based on measurements of electrical resistance and was supported by experiments performed with a short-circuited coil and rings carrying the superconducting current (see \cite{Delft} for references). In particular, by results of  a remarkable Onnes' experiment  reported by Willem Keesom at the 4th Solvay conference in 1924\footnote{In this experiment a superconducting lead sample (either a ring or a spherical shell) with the field-induced persistent current was suspended on a torsion spring in a horizontal magnetic field produced by a fixed superconducting ring concentric with the suspended sample. Originally the magnetic moment induced in the sample was inline with the field. Then the spring with the sample was turned for  30$^{^o}$. It was expected that the transverse Lorentz force acting on the superconducting current carriers in the sample will tend to decrease the angle. However, the angle stayed undiminished over more than 6 hours of observation for each sample. Onnes concluded that (a) the upper limit of the ratio of resistivity in the S state to that in the N state (at $T$ slightly upper $T_c$) is less than $10^{-12}$; and (b) probably, the  transverse Lorentz force does not act on the superconducting charge carriers and therefore  the Hall effect in superconductors is absent.}  \cite{Mehra,Onnes-1924}.

Recall that the hallmark of  perfect conductors is the irreversibility of their magnetic properties, which means that thermodynamics is inapplicable to describe the properties of superconductors \cite{London50,Shoenberg,VK}. In its turn, this implies that there is no phase transition  at the S/N transition. This picture follows from the Maxwell electrodynamics as applied to resistanceless samples; it seemed so obvious that it looked pointless to test it \cite{Shoenberg}.

However, the jump in heat capacity discovered by Keesom with collaborators indicated that this 
picture may actually be incorrect. Namely, that the S/N transition can represent a phase transition associated with yet unknown alteration of the electron structure, as it has been suggested in number of occasions before\footnote{E.g., at the 1st Solvay conference in 1911 Langevin suggested that superconductivity can be associated with a new diamagnetic state. Similar proposal was sounded by Langevin and Bridgman at the 4th Solvay conference in 1924 to explain the existence of the temperature dependent critical field $H_c(T)$. However, Keesom and Lorentz did not think that thermodynamics could be applicable in this case \cite{Mehra}.  }.

Keesom's results served as a powerful call to study magnetic properties and corresponding experiments were set up in Berlin, Kharkov, Toronto and Oxford. The first convincing results were obtained in Berlin (Meissner's group) \cite{Meissner} and Kharkov (Shubnikov's group) \cite{Shubnikov}. These historical experiments are briefly considered below. Details can be found in \cite{Wilhelm}. 

\textbf{Meissner and Ochsenfeld} \cite{Meissner} reported on measurements of the magnetic field in the vicinity and inside superconducting samples in four arrangements. The samples were single-crystalline tin and poly-crystalline lead cylinders (130-140 mm in length and 10 mm in diameter \cite{Wilhelm}) placed vertically in the  horizontally applied uniform magnetic field $\textbf{H}_0$. The measurements were performed using a small search coil connected to a ballistic galvanometer. The coil could be moved round the sample and \tg{rotate without opening} the cryostat. The current induced at turning the coil for 180$^{^o}$ is proportional to the coil cross-sectional area and the induction $B$ in the coil location, thus allowing to find an averaged field over the coil volume. As the main sources of error, the authors indicate insufficiently accurate knowledge the spacial distribution of the coil winding and the imperfection of the cylindrical shape of the samples, especially of the single-crystalline tin. 

In the first arrangement the field distribution near one sample was measured after it was cooled below $T_c$ in  $H_0\approx$ 5 G. According to the Faraday law, the field should stay undisturbed since there is no e.m.f. induced and the magnetic permeability of the sample materials $\mu_m$ negligibly differs from unity. However, it turned out that below $T_c$ the field pattern near the sample changed almost to that which would be expected if $\mu_m$ of the superconductor is  zero or the magnetic susceptibility $\chi=-1/4\pi$ (the authors used cgs units, which will be also used throughout this paper). 

In the second arrangement two parallel either tin or lead samples were cooled in the same transversely applied field. It was found that below $T_c$ the field between the tin samples increased for a factor 1.70; for the lead samples this factor was 1.77. The increase factor for the field in the location of the search coil  calculated coming from zero permeability of the S state was 1.77. These data support the statement above about zero permeability of the sample material in the S state. Note that no eddy currents can be induced under conditions of these experiments unless the Faraday \tg{Law} is broken.  
   
In the third arrangement the sample was a hollow lead cylinder (a tube with the wall 2 mm thick). It was again cooled through $T_c$ in the same field as before and $B$ was measured inside the tube and adjacent to it outside. It was found that the outer field changed in about the same way as it was for the solid cylinder. The inner field changed also: it increased for about 5\% (Smith and Wilhelm \cite{Wilhelm} discussing this experiment in 1935 name this increase as 10\%). The authors were not able to establish if the field inside remained uniform. On switching off the applied field keeping the sample superconducting, the field inside remained unchanged; at the same time the field outside decreased but it did not become zero. 

In regard of this arrangement, the authors noted that their observations may look inconsistent with the statement about zero permeability. They suggested that these results can be explained in terms of microscopic or macroscopic currents in the superconductor assuming that $\mu_m=1$ for the current-free regions. Now we understand (see, e.g. \cite{VK}) that due to a non-ellipsoidal shape of the tube sample it was in neither one of the equilibrium superconducting states, the field passes through it via irregular N domains, and the flux is trapped when $H_0$ is switched off. One can add that the observed field enhancement inside the tube sample is consistent with recent direct measurements of the field near the sample in the intermediate state \cite{IS-3}. This means that there is no contradiction between results of the first two arrangements with those of the third one, and that Meissner and Ochsenfeld were exactly right in their interpretation.

After all, in the fourth arrangement two tin samples used in the second arrangement were  connected end-to-end in series and a dc current of about 5 A was introduced through their other ends.  As found, the field between the samples was greater below $T_c$ than that above it, although the current was kept unchanged. Smith and Wilhelm noted that the measured field was about the same regardless whether the current was introduced before or after the sample passed through
$T_c$, and in both cases the field readings were greater than that calculated assuming the surface superconducting current. This observation is largely ignored in textbooks; we will come back to it later. 

The main conclusion of Meissner and Ochsenfeld was that $B$ in the S phase is always zero, however not all researchers agreed with that (see, e.g. \cite{Mendelson_1934}). The experiment of Rjabinin and Shubnikov removed  all doubts.   

\textbf{Rjabinin and Shubnikov} \cite{Shubnikov} attacked the same problem via measuring the magnetic moment $\textbf{M}$ of a superconducting lead rod (5 mm in diameter and 50 mm long) at constant temperature 4.2 K vs $\textbf{H}_0$ applied parallel to the sample longitudinal axis.  Two methods were used, which are similar to those employed in contemporary ac and dc magnetometry. In the first method the change $\Delta \textbf{M}$  was determined by measuring the current induced in a pickup coil tightly wound around the middle of motionless sample at a sadden change of the applied field in small steps $\Delta \textbf{H}_0$.  In the second method the current in the pickup coil was induced by quickly removing the sample away from the coil without changing $\textbf{H}_0$. Results obtained by both methods were consistent with each other but the second method appeared to be more reliable. So, the discussion was mainly based  on the results obtained via dc measurements. The reported data were $\textbf{B}$ vs the field intensity $\textbf{H}$ inside the sample, which in the chosen geometry  equals $\textbf{H}_0$. 

It was found that (a) when the sample was first magnetized (i.e. after cooling in zero applied field) $B$ and $\mu_m$ were zero at $H\leqslant H_c$; in a narrow field interval near $H_c$ the induction rapidly changed to a magnitude equal to that in the normal metal; at the $H>H_c$ $\mu_m=1$.
(b) At decreasing $H$,  $B = H$ until $H$ reached its critical value; at $H$ close to $H_c$, the induction experienced a sudden jump down, but it did not become zero; with a further decrease of the field $B$ was also decreasing; at $H=0$ there  remained a residual induction close to 18\% of the maximum $B$ at $H=H_c$. The observed in such a way magnetization \textit{loop} was reproducible.  

The authors concluded that ``the actual fact that a jump takes place in the induction in falling field strengths we are incline to ascribe to the formation of a new phase with $B=0$." The incomplete reversibility of the data obtained was attributed to imperfections of the sample material. As now well known (see, e.g. Fig.\,3 below) the interpretation of Rjabinin and Shubnikov was correct. 

The experimental results of Meissner and Ochsenfeld plus Rjabinin and Shubnikov, confirmed in experiments of Tarr and Wilhelm \cite{Tarr}, and Mendelssohn and Babbitt \cite{Mendelson_1935}  once and for all changed the landscape of superconductivity. Specifically, it was established that at definite conditions a superconducting sample can be found in a reversible state, referred to as the Meissner state (MS), which is characterized by zero induction simultaneously with zero resistivity. This is the essence of the Meissner effect. However, paradoxical as it may sound, the physics of this well known phenomenon, as shown below, still remains an unsolved problem. To discuss a possible way to resolve it is the main objective of this review. 

\section{MEISSNER STATE DEFINITION}  
Before discussing the theories, we have to specify the definition of the MS.

It is defined as a thermodynamic (and therefore reversible) state at which the induction $\textbf{B}$  throughout  the volume $V$ of a massive superconducting  body (sample) placed in a static  magnetic field $\textbf{H}_0$ is zero.

The MS can be also defined as the S state at which the magnetic susceptibility $\chi$ all over the volume of the massive body is $-1/4\pi$.  

The massive body is considered to be as such if its dimensions greatly 
exceed a so-called penetration depth $\lambda$, i.e.,  the width of a near-surface layer within which the induction of the external field near the sample $\textbf{B}_{ext}(=\textbf{H}_{ext}$ in cgs units)\footnote{In general  $\textbf{H}_{ext}\neq \textbf{H}_0$ but it is always parallel to the surface of the sample in the MS due to continuity of the normal component of $\textbf{B}$ at the sample boundary, i.e. the external field bends around the sample.} decays down to zero inside it.

The MS is observed only in sufficiently pure (see footnote $(^2)$ on p.\,82 in \cite{VK}) 
singly connected samples of an ellipsoidal shape in a range of the applied field $0< H_0 < H_{c1}(1-\eta)$, where $\eta$ is a demagnetizing factor with respect to the sample axis parallel to $\textbf{H}_0$, and $H_{c1}$ is the lower critical field of type-II superconductors. In type-I materials $H_{c1}=H_c$, where  $H_c$ is the thermodynamic critical field. The latter is a measure of the condensation energy $E_c$ defined through the relationship $E_c=(H_c^2/8\pi)V$. Recall that $\eta$, the proportionality coefficient between the demagnetizing field $\textbf{H} _d$ and magnetization $\textbf{I}$, is well defined only for ellipsoidal bodies with uniform $\textbf{I}$ \cite{Maxwell,Landafshitz_II,VK}.

If $\textbf{H}_0$ is not parallel to either one of the ellipsoidal axes,  it should be broken for  components parallel to the axes and $\eta$ in the formula for the MS field range is the maximum demagnetizing factor for the given sample-field configuration. In particular, for a planar sample in a non-parallel field, the maximum $\eta$ equals one\footnote{Strictly speaking, a phrase like "superconducting samples with $\eta=1$" has no sense because $I$ in this case in not uniform and, therefore, $\eta$ is undefined. As was first shown by Maxwell, superconductors with $\eta=1$ do not exist \cite{Maxwell,VK}. The above phrase should be understood as a superconducting infinite plate in a perpendicular field.} and therefore such a sample does not exhibit the MS in any $\textbf{H}_0$, regardless how small this field is \cite{VK}.

On the other hand, a sample with $\eta=0$ (referred to as the sample of cylindrical geometry) is in the MS at $H_0< H_{c1}(=H_c$ for type-I materials). This can be a long cylinder (not-necessarily a circular one), an infinite slab or a wide ribbon-like foil in the field parallel to its generating  line  \cite{VK}. In all such cases the applied field $\textbf{H}_0$ stays undisturbed all the way down to the sample surface, i.e. $\textbf{H}_{ext}=\textbf{H}_0$.  

Other (inhomogeneous) equilibrium S states, such as the intermediate and mixed states in type-I and type-II materials, respectively, are also observed only in samples of the ellipsoidal shape\footnote{This can be easily understood from consideration of a magnetization curve of a sample in thermodynamic equilibrium. The inhomogeneous state occupies an upper part of this curve, whereas its lower part (the one at low field) is taken by the MS. So, if a non-ellipsoidal sample can be in equilibrium in the inhomogeneous state, i.e. at the high field, it should be in equilibrium at the low field as well. This implies that the non-ellipsoidal sample can  be in the MS, which has never been observed.} \cite{Shoenberg}. A common feature of the ellipsoidal samples in either homogeneous or inhomogeneous states is uniformity of the field intensity (the field strength or, as Maxwell names it \cite{Maxwell}, the magnetizing force) $\textbf{H}$ throughout their volume \cite{VK}.  

In non-ellipsoidal samples $\textbf{H}$ is not uniform and such samples can not be entirely in either one of the equilibrium states\footnote{Some symmetrical but not ellipsoidal bodies in the field parallel to the symmetry axis can mimic the MS in a sense that their magnetization curve at low $H_0$ can be linear. However, unlike the bodies in the genuine MS, their average susceptibility $\chi$ differs from $-1/4\pi$.}.  Therefore, the MS can be also defined as the equilibrium S state in which the field intensity $\textbf{H}$ with magnitude less than $H_{c1}$ is uniform throughout the volume of the massive  body. 

All three definitions of the MS are identical, i.e. each one unambiguously follows from the other. 

Finally, let us pay attention to one more important circumstance\footnote{The author is grateful to professor Kresin for pointing out this moment.} associated with the fact that any equilibrium system (whether it is classical or quantum)  have to possess symmetry  with respect to reversal of time \cite{LL-QM}. This implies that if currents are present in the equilibrium state, as it takes place in the MS, they must mutually compensate each other so that a total current does not arise\footnote{Naturally, this does not apply to multiply connected bodies since they cannot be in equilibrium in a magnetic field.} \cite{Landafshitz_II}. Evidently, this rule is equally related to the intermedium and the mixed states.

\section{TWO-FLUID MODEL}

The two-fluid model of Gorter and Casimir \cite{Gorter_Casimir} (see also \cite{Shoenberg,Kes_2012,Wilhelm}) is a thermodynamic theory addressing properties of superconductors in zero field. Its key idea is that the conduction electrons of the superconducting material are divided for two interpenetrating groups or fractions with different energy levels. This turned out very fruitful idea is used in all theories of superconductivity and superfluidity ever since. The fraction $x$ with the higher (Fermi) energy represents  "non-condensed" or "normal" electrons, correspondingly another fraction (1-$x$) represents "condensed" or "superconducting" electrons. As postulated, properties of the latter fraction are characterized by zero entropy, which means that the superconducting electrons are supposed to be completely ordered. This postulate is based on the experimental fact of the absence of thermoelectric effects in superconductors \cite{Shoenberg,Wilhelm}. The fractions are functions of temperature: $x=0$ at $T=0$, and  $x=1$ at $T=T_c$. 

The free energy\footnote{Since the system (sample) is in zero field, there is no difference between the Helmholtz, Gibbs and total free energies. On the same reason the free energy does not depend on the sample shape. Since the sample is not magnetized,  the free energy density can be used regardless of the sample shape \cite{VK}.} of electrons per unit volume $f(T)$ is chosen as
\begin{equation}
	f(T)=x^{0.5}f_n(T) +(1-x)f_s(T),
\end{equation} 
where subscripts $n$ and $s$ designate the N and S fractions, respectively.  

\tb{The free energy density of the N fraction is chosen as $f_n=-\alpha T^2/2$ to fit the liner dependence of the electron heat capacity in normal metals. The power ${0.5}$ in the first term is chosen to fit an observed quadratic temperature dependence of the thermodynamic critical field $H_c(T)$ which, as it was shown by Kok \cite{Kok_34}, images the cubic temperature dependence of the electron heat capacity in superconductors.} The free energy density of the S fraction is chosen as $f_s=-\beta=const$, what reflects the zero entropy postulate. The coefficients $\alpha$ and $\beta$ are parameters characterizing properties of the N and S fractions, respectively.   

At equilibrium $(\partial f/\partial x)_T=0$. From that with the use of condition $x(T_c)=1$ it follows that 
\begin{equation*}
 x = \left(\frac{T}{T_c}\right)^4 \tag{1a}
\end{equation*} 
and 
\begin{equation*}
	\beta=\frac{\alpha T_c^2}{4}.\tag{1b}
\end{equation*}

After substituting Eqs.\,(1a) and (1b) into Eq.\,(1) and using the same condition $x(T_c)=1$, \tg{Eq.\,(1)} takes form
\begin{equation*}
	f=-a\frac{T_c^2}{4}-a\frac{T^4}{4T_c^2}.
\end{equation*}

Therefore, the electron specific entropy is
\begin{equation*}
	s=-\frac{d f}{d T}=\alpha\frac{T^3}{T_c^2}.\tag{1c}
\end{equation*}

Hence, in spite of zero entropy of the superconducting electrons, the entropy of the S fraction is not zero due to the temperature dependence of $(1-x)$, which is a relative number density of the superconducting electrons in the London theory. 

Next, the specific heat capacity is
\begin{equation*}
	c=T\frac{ds}{dT}=3\alpha\frac{T^3}{T_c^2}.\tag{1d}
\end{equation*}

\tb{This justifies the chose of the power ${0.5}$ in Eq.\,(1).}

In a few more steps (see, e.g. \cite{Shoenberg}) the two fluid model yields
\begin{equation*}
	\beta=\frac{H_{c0}^2}{8\pi},\tag{1e}
\end{equation*}   
where $H_{c0}$ is the critical field at $T=0$. 

Therefore, $\beta$ equals the condensation energy density at $T=0$, which implies that the second term in Eq.\,(1) represents a temperature dependent difference (gap) of the energies of electrons in the N and S fractions, similar as it was later found in the theory of Bardin, Cooper and Schrieffer (BCS)
 \cite{Schrieffer,BCS}.

The Keesom formula for the latent heat and the Rutgers formula for the specific heat difference at the S/N transition (see, e.g., \cite{VK}) can also be derived from the two-fluid model. All formulae \tb{of the model} nicely fit experimental data. It is worth reminding that it was the two-fluid model (specifically Eq.\,(1a)) that provided the success of the London \tg{theory, and} Eq.\,(1a) itself is a demonstration of the predicting power of thermodynamics \tb{at an  adequately chosen thermodynamic potential. }

Overall, so broad list of successful formulae leaves no room to question the correctness of the both fundamental assumptions of the two-fluid model, namely of the assumption about the interpenetrating fluids and of the zero-entropy postulate. Hence, there are three ``big zeroes" characterizing superconductivity, i.e. properties of the superconducting electrons: zero resistivity, zero induction and zero entropy. Respectively, a theory which does not lead automatically to all these three zeros cannot be complete. 
In fact, the incompleteness is the only disadvantage of the two-fluid model, but it has never pretended to be considered as a complete theory.

\section{LONDON THEORY} 

A description of the electromagnetic properties of superconductors in the MS is given in the theory of  Fritz and Heinz London \cite{London35,London50}; with modifications it is adopted in the theory of  Ginzburg and Landau (GL) \cite{GL} and in the BCS theory \cite{BCS,Schrieffer}. 
 
The London theory is based on  two equations. The first one follows from Newton's acceleration equation for a free and spinless electron, which is assumed to be applicable to  the superconducting electrons, i.e to the charge carriers not experiencing resistance. This equation reads  
\begin{equation}
	m\dot{\textbf{v}}=e\textbf{E},
\end{equation}
where $m$ and $e$ are the mass and charge of one of these electrons, respectively, $\dot{\textbf{v}}$ is the time derivative of its velocity, and $\textbf{E}$ is the electric field acting on it. 
	
Coming from Eq.\,(2), the time derivative of the current density $\textbf{j}(=n_se\textbf{v}$, where $n_s$ is the number density of the superconducting electrons) is  	     
\begin{equation*}
	\dot{\textbf{j}}=\frac{c^2}{4\pi\lambda_L^2}\textbf{E}, \tag{2a}
\end{equation*}

This is the first London equation, where $c$ is the electromagnetic constant of the cgs unit system equal to the speed of light and $\lambda_L$ is a temperature dependent material constant referred to as the London penetration depth, which is defined as\footnote{Note that if the superconducting electrons are combined in Cooper pairs $\lambda_L$ remains unchanged.}
\begin{equation}
	\lambda_L=\left(\frac{mc^2}{4\pi n_s e^2}\right)^{1/2}.
\end{equation}  

The second London equation is 
\begin{equation}
\nabla\times \textbf{j}=-\frac{c}{4\pi \lambda_L^2}\textbf{H},
\end{equation}
where $\textbf{H}$ is the magnetic field intensity acting on the superconducting electrons  \cite{London35}. Important to stress that this is \textit{not} the induction $\textbf{B}$, as written in some textbooks\footnote{Recall	that $\textbf{H}$ is a magnetic field acting inside a body  on  a ``native" (i.e. belonging to the body) charge. $\textbf{H}$ does not include the field due to the charge in question, whereas the induction $\textbf{B}$ (the average microscopic  field caused by all native charges) includes it. On that reason a ``foreign" (probing) charge always experiences the action of $\textbf{B}$, but not $\textbf{H}$ \cite{VK}.}.    

Eq.\,(4)  is the   famous London equation for the current density in superconductors replacing the Ohm Law in normal metals. In \cite{London35} it is derived from Eq.\,(2a) 
taking curls from its both sides and using the Maxwell equation $c\nabla\times \textbf{E}=-\partial \textbf{B}/\partial T$ along with a set of assumptions. Specifically,  
\begin{equation}
\mu_m=\varepsilon_e=1,  
\end{equation} 
where  $\mu_m$ and $\varepsilon_e$ are the magnetic permeability  and dielectric permittivity of the superconducting material, respectively. 

In \cite{London35} the assumptions Eq.\,(5), equating electromagnetic properties of superconductors with those of vacuum, are introduced as a simplification. The authors explain that "we do not know anything about it", adding that "This may subsequently to be corrected". However, 15 years later F. London \cite{London50} keeps these assumptions in force and on the best of our knowledge they have never been reconsidered afterwards.

Two other assumptions are \cite{London35,Shoenberg}   
\begin{equation}
\textbf{B}_\infty=\textbf{j}_\infty=0,
\end{equation}
where subscript $\infty$ designates the quantities in the sample interior or at the depth much greater than $\lambda_L$.

The first of these two assumptions is a postulate based on the Meissner effect, and the second one is justified by a never published theorem ascribed to Bloch. F. London formulates it as ``in the absence of an external field the most stable state of any electronic system is a state of zero current" \cite{London50}. Hence, due to the absence of the field within the sample in the London theory, the second assumption in Eq.\,(6) follows. Note that $\textbf{j}_\infty=0$ directly stems from the time reversal rule mentioned above.     

The assumptions of Eq.\,(6) lead to 
\begin{equation*}
	\textbf{H}_\infty=\textbf{v}_\infty=0.\tag{6a}
\end{equation*}

Here $\textbf{H}_\infty=0$ because  $\mu_n=1$ (Eq.\,(5)) and $\textbf{B}_\infty=0$ (Eq.\,(6)); and  $\textbf{v}_\infty=0$ because $\textbf{j}_\infty=0$ (Eq.\,(6)). 

From Eq.\,(2a) using of the Maxwell equations $c\nabla\times \textbf{B}=4\pi \textbf{j}$ and $\nabla \cdot \textbf{B}=0$ along with conditions $\mu_m=1$ (Eq.\,(5)) and $H_\infty=0$ (Eq.\,(6a)) one obtains (see, e.g., \cite{Shoenberg})
\begin{equation}
\lambda_L^2\nabla^2\textbf{H}=\textbf{H}.
\end{equation}  

Analogically, with the use $\textbf{j}_\infty = 0$ and taking into account that no external current is fed into the sample (i.e. $\nabla\cdot \textbf{j}=0$) one obtains 
\begin{equation}
\lambda_L^2\nabla^2\textbf{j}=\textbf{j}.
\end{equation}  
 
Eqs.\,(7) and (8) imply that in the massive samples the  induction $\textbf{B}(=\textbf{H}$ in the theory) and the current density $\textbf{j}$ exponentially decay with depth from their values at the sample boundary to zero in its interior with the decay constant $\lambda_L$ in both cases. Hence, taking into account the smallness of $\lambda_L(\sim 10^{-6}$ cm assuming one superconducting electron per atom), it looks as  the theory meets the Meissner condition (zero $B$ in the volume of the massive sample) explaining it as a screening effect of the current (\textit{assumed} circumferential\footnote{The idea of the field induced circumferential surface current appeared well before the Meissner-Ochsenfeld effect was discovered and the London theory proposed. E.g., it was used by Lorentz \cite{Lorentz} for interpretation of Onnes' experiment with superconducting shell (see footnote ($^1$) above). Hall challenged this idea \cite{Hall}. Nowadays the assumption that the induced current is the circumferential one is taken for granted. However, as we will see, this is one of the most controversial assumptions of the standard theories. Note also, that it conflicts with the time reversal rule.}) persistently running in a thin surface layer.

So far, the theory may appear as a rather strange construction\footnote{In the first edition of Shoenberg's book \cite{Shoenberg} (1938) the London theory was hardly mentioned. Pippard recollects, that when he asked why, Shoenberg said that Landau thought it ill-founded \cite{Pippard_2005}. For reference: Shoenberg wrote this book being in Kapitza's institute in Moscow, where Landau transferred from Kharkov Physical-Technical Institute after arrest of Shubnikov in 1937. Shubnikov was executed in the same year.  In 1938 Landau was also arrested. In one year he was released owing to an extraordinary action of Kapitza (see, e.g., \cite{Shoenberg-Kapitza}).}, the sole purpose of which is to fulfill deliberately introduced (Eq.\,(6)) the Meissner condition $B=0$. Therefore, it was quite unexpectedly when it turned out that, being combined with the two-fluid model\footnote{As mentioned above, $n_s/n_{s0}$ in the London theory corresponds to $(1-x)$ in the two-fluid model. Accordingly,  $n_s/n_{s0}=1-(T/T_c)^4$, where $n_{s0}$ is $n_s$ at $T=0$.}, results of the London theory appeared to be consistent with experimental  data on the temperature dependence of the penetration depth \cite{Shoenberg}. 
  
Another form of the London equation (4) is 
\begin{equation}
\textbf{j}=-\frac{c}{4\pi \lambda_L^2}\textbf{A},
\end{equation}
where $\textbf{A}$ is the vector potential defined as $\textbf{H}=\textbf{B}=\nabla\times \textbf{A}$.

To restrict multiplicity of $\textbf{A}$, the theory imposes  supplementary conditions\footnote{Eq.\,(10) is the standard supplementary condition for the vector potential. Physics behind it is that the current cannot pass through the metal/insulator interface \cite{Tamm}. In quantum mechanics Eq.\,(10) provides the commutativity of operators of the generalized linear momentum and of the vector potential \cite{LL-QM}. }     
\begin{equation}
\nabla\cdot \textbf{A}=0,
\end{equation}
and 
\begin{equation}
\textbf{A}_{\perp}=0,
\end{equation}
where $\textbf{A}_{\perp}$ is a component of the vector  potential inside the sample normal to its surface. 

Given Eq.\,(9), the condition Eq.\,(10) reflects the continuity of the induced current (no source, no sink), implying that the current and the lines of the vector potential make closed loops. 
Eq.\,(11) implies that the current cannot cross the sample surface ($\textbf{j}_{\perp}=0$). 

Eq.\,(9) is that form of Eq.\,(4) for which  Pippard suggested a non-local extension of the London theory  \cite{Pippard53}. It is also \tg{consistent} with experiment \cite{Andreas,NL_neutrons,NL}.

After all, Eq.\,(9) can be written (substituting $\textbf{j}=en_s\textbf{v}$ and  $\lambda_L$ given in Eq.\,(3)) in a microscopic form as
\begin{equation}
m\textbf{v}+\frac{e}{c}\textbf{A}\equiv\widetilde{\textbf{p}}=0,
\end{equation}
where $\widetilde{\textbf{p}}$ is a generalized linear  momentum \cite{LL-field} of a single superconducting electron. 

Eq.\,(12)  is highly remarkable, as it was shown by F. London himself \cite{London50}.  He obtained Eq.\,(12) in a different way and showed that Eq.\,(4) follows from Eq.\,(12). Hence, the theory can be constructed starting from Eq.\,(12) as a postulate. At the same time he did not reconsider the assumptions (5) and (6). Note, that due to Eq.\,(9) $\textbf{j}_\infty=0$ implies that $\textbf{\textbf{A}}_\infty=0$ as well.

Here are some consequences of Eq.\,(12).

1. The London equation (4)/(9) takes an extremely simple form\footnote{Simplicity is the doubtless merit of any theory. Citing Feynman, ``Nature has a great simplicity and therefore a great beauty"\cite{Feynman_Phys Laws}. }:  $\widetilde{\textbf{p}}=0$. 

2. Since in magnetic field de Broglie's wave length is $\lambda_{DB}=h/\widetilde{p}$, Eq.\,(12) implies that $\lambda_
{DB}$ of superconducting electrons is infinite, indicating that the superconducting body is a macroscopic quantum object.  

3. Since the London theory is not restricted by any particular area of the phase diagram of the MS, Eq.\,(12) should be valid regardless on parameters  of this state (e.g., temperature $T$ and the applied field $H_0$). On this ground F. London concluded that the S \textit{phase}\footnote{The S phase is defined as a phase of a superconducting material in which $n_s\neq 0$. The S phase can occupy the entire sample volume (as it takes place in the MS) or a part of it.} possess a long-range order, characterized by  the "rigid" (i.e., independent on the state parameters) linear momentum $\widetilde{p}(=0)$. F. London foreseen that this rigidity is a cornerstone of superconductivity,  which ``offers a remarkable possibility of reducing superconductivity to an apparently very simple model" \cite{London50}.

4. For multiply connected bodies the concept of rigidity led F. London to prediction of the magnetic flux quantization in superconductors \cite{London50}. A decade later the latter was confirmed experimentally \cite{Deaver}, albeit with a factor 1/2 absent in the original prediction; this factor, as it was shown by Onsager \cite{Onsager}, stems from the pairing concept of the BCS theory \cite{Schrieffer}. A pure theoretical prediction of the flux quantization is, in any respect, the highest achievement of the London theory.

To summarize, we see that the London theory has an impressive list of achievements and 
it cannot be denied that this theory served as the basis for understanding superconductivity \cite{Gorter64,Shoenberg}. However, it is hard not noticing  serious difficulties in  this theory\footnote{From the very beginning (see \cite{London35}) the Londons well understood the weakness of the theoretical background of Eqs.\,(2a) and (4). In \cite{London50} F. London does not derive them stating instead that the validity of these equations follows from the experimental confirmation of the consequences which they imply.}. Let us look closer at some of them. But before that we remind two facts solidly established in experiments (see, e.g., \cite{Shubnikov,Desirant-Shoenberg,MS}) and justified by thermodynamics \cite{VK}.
\begin{figure}
	\includegraphics[width=0.5\linewidth]{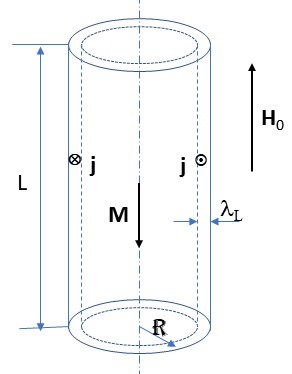}
	\caption{A massive cylindrical sample in the London theory. $L\gg\Re\gg\lambda_L$, the demagnetizing factor $\eta=0$. The London penetration depth $\lambda_L$ is an effective width of a surface layer containing the field induced circumferential screening current. The   induction $\textbf{B}$ and the current density $\textbf{j}$ in this layer equal $\textbf{B}_0=\textbf{H}_0$ and  $c\textbf{n}\times \textbf{H}_0/4\pi\lambda_L$, respectively; behind this layer $\textbf{B}$, $\textbf{H}$, $\textbf{A}$, and $\textbf{j}$ are zero. $\textbf{M}$ is the sample magnetic moment caused by the surface current. $\textbf{H}_0$ is the applied field. }
	\label{fig:epsart}
\end{figure} 

(a) The magnetic moment of a sample in the MS [the sample has the ellipsoidal shape 
and $\textbf{H}_0$ is parallel to one of its axes\footnote{At another orientation of $\textbf{H}_0$, it should be broken for components parallel to the sample axes and $\textbf{M}$ is the vector sum of the moments relative to each axis; in such case $\textbf{M}$ can be not aligned with (antiparallel to) $\textbf{H}_0$, but it is always antiparallel to $\textbf{H}$. }, demagnetizing factor with respect to which is $\eta(\neq 1)$] is 
\begin{equation}
\textbf{M}=-\frac{V}{4\pi(1-\eta)}\textbf{H}_0.
\end{equation}

(b) The magnetic energy of the sample in the MS is 
\begin{equation}
E_m\equiv-\int \textbf{M}\cdot d\textbf{H}_0=-\frac{\textbf{M}\textbf{H}_0}{2}=\dfrac{VH_0^2}{8\pi (1-\eta)}.
\end{equation}

In magnetostatics,  $E_m$ represents the work done by the magnetic field (executed by an electric field generated while the magnetic field is changing) to magnetize the sample. By virtue of the energy conservation, it is equal to the variation of the kinetic energy of electrons $\Delta T$ plus the energy of the outer field produced by the magnetized sample. In diamagnetics $\Delta T=E_k$, the latter being the kinetic energy of the field-induced motion of electrons. In  diamagnetic samples of the cylindrical geometry $E_m=E_k$ because the outside field caused by the sample is absent \cite{VK,Tamm}.

A schematic of the massive sample in the MS in the London theory is shown in Fig.\,1\footnote{In \cite{London50} and \cite{Shoenberg} solutions of the London equations for the field, current and the magnetic moment are available for cylinders, plates and spheres. The calculated magnetic moment of the cylinder and sphere with radius $\Re\gg\lambda_L$ equals the moment of these figures with zero induction and the radius $\Re-\lambda_L$; for the plate of thickness $d\gg\lambda_L$ the moment equals that of the plate with $B=0$ having the thickness $d-2\lambda_L$.}. An ``active" part of this sample (i.e. the part where $\textbf{B}$ and $\textbf{j}$ are not zero) is a surface layer  with an effective width\footnote{The effective width of the penetration layer is defined as $\lambda_{eff}\equiv H_0^{-1}\int_0^\infty B(z)dz$, where $z$ is the depth from the surface and $B(z)$ is the depth's profile of the induction. Inside this layer $B=B_{eff}\equiv(\lambda_{eff})^{-1}\int_0^\infty B(z)dz$ and $j=j_{eff}\equiv(\lambda_{eff})^{-1}\int_0^\infty j(z)dz$, where $j(z)$ is the depth's profile of the current. In the cylindrical sample of the London theory $\lambda_{eff}=\lambda_L$, $B_{eff}=H_0$ and $j_{eff}=g/\lambda_L=cH_0/4\pi\lambda_L$, where 
$g$ is the linear current density.} $\lambda_L\ll\Re$. Behind this layer all magnetic characteristics  are zero (Eq.\,(6)), implying that the sample interior, nearly whole its volume, is totally inert. If so, there should be no difference if the interior is in the S or in the N state, or there is no interior at all. 
\begin{figure}
	\includegraphics[width=0.5\linewidth]{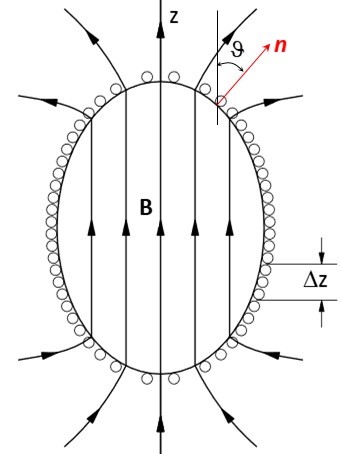}
	\caption{A current shell having a form of an ellipsoid of revolution (spheroid) with respect to the vertical axis $z$; $\vartheta$ is the angle between $z$ and the unit vector $\textbf{n}$ normal to the shell. If the current linear density $g$ is proportional to $\sin \vartheta$ or, equivalently, there are equal currents in each axial interval $\Delta z$, the magnetic field inside this shell $\textbf{B}(=\textbf{H}$ since $\mu_m=1)$ is uniform and parallel to $z$.} 
	\label{fig:epsart}
\end{figure} 

Thus, in the London theory magnetic properties of solid (continuous) and hollow samples are identical. However, as is well known \cite{Maxwell,Jackson}, properties of the solid and hollow magnetized bodies are different. As we know, for superconductors it was clearly demonstrated by Meissner and Ochsenfeld. This can be also seen from the fact that $M$ and $E_m$ (Eqs.\,(13) and (14)) are proportional to the volume of the sample, like for all other continuous bodies where $M$ and $E_m$ are caused by magnetization, and include parameters of neither the cavity nor the wall.    
   
More specifically, in the London theory a long circular cylindrical sample in  the longitudinal field $\textbf{H}_0$ is identical to a solenoid of the same shape and in the same field with the current per unit length  $\textbf{g}=c\textbf{n}\times \textbf{H}_0/4\pi$\footnote{This current is calculated from the boundary condition for the tangential component of the induction $B_t$, which stems from the condition of continuity for $H_t$ \cite{VK,Landafshitz_II}.}, where $\textbf{n}$ is the unit vector normal to the surface and directed outward; the moment $\textbf{M}$ produced by this current is the same as that in Eq.\,(13) with $\eta=0$ (see problem 1.4 in \cite{VK}). The field inside the solenoid due to this current equals $-\textbf{H}_0$, it compensates the field $\textbf{H}_0$ resulting that the field there (both $\textbf{H}$ and $\textbf{B}$ since the solenoid is empty and therefore $\mu_m=1$) is zero. Hence, the \textit{solenoid's interior} is screened from the applied field, or the field is expelled from the solenoid when the current is turned on. 

On the other hand, the long solenoid is a particular case of  a current shell shaped as an ellipsoid of revolution also referred to as a spheroid. As known (see, e.g., \cite{Feynman_Lectures}), the field inside such a shell is uniform provided $g\sim \sin\vartheta$ (see Fig.\,2). Then in an axial field $\textbf{H}_0$ the field inside the shell is zero  if  $\textbf{g}=c\textbf{n}\times \textbf{H}_0/4\pi(1-\eta)$, where $\eta$ is the demagnetizing factor of a solid spheroid of the same shape as the shell (see, e.g., problem 1.7 in \cite{VK}). Therefore, the above reasoning about the screening current and the expelled field in the solenoid holds for  the spheroidal current shell. The moment $\textbf{M}$ of such a  shell is the same as that in Eq.\,(13) (see problem 1.8 in \cite{VK}). 

On the contrary, inside samples in the MS the induction $\textbf{B}=0$, but the field intensity $\textbf{H}$ is not. In the cylindrical sample, e. g., like that shown in Fig.\,1,  $\textbf{H}=\textbf{H}_0$\footnote{This follows from the Poisson theorem  and in this case also from the continuity of  $H_t$ \cite{VK,Landafshitz_II}.}. However, someone  can  (perhaps) say that the field due to the induced circumferential surface current compensates the field $\textbf{H}$ resulting in zero field (both $B$ and $H$ as stated in the London theory) inside the sample\footnote{A statement of such kind can be found in some textbooks, however it 
is incorrect because it contradicts to the boundary condition for $H_t$ (continuity) which is used to calculate $\textbf{g}$ \cite{VK,Landafshitz_II}.}.

Now, taking into account what was said about the spheroidal current shell, can one say that the assumption of the induced circumferential current provides a consistent picture of the MS? The answer is \textit{no} already because the MS is observed in samples of any ellipsoidal shape,  but not only in the ellipsoids of revolution. For example, it can be the same cylinder as that in Fig.\,1 but in the transverse field \cite{Shoenberg} or a film in the  parallel field, like the film which magnetization curves are shown in Fig.\,3. Another reason of this negative answer consists in the fact that the field inside the spheroidal sample, produced by the circumferential current needed to obtain correct magnetic moment, equals $-\textbf{H}_0$, while the field which is supposed  to be compensated is $\textbf{H}_0/(1-\eta)$ (see problem 1.7 in \cite{VK}).

Another difficulty of the London theory can be seen from the following.  A sample with $\eta=0$ can be an infinite plate in the parallel field, like the film in Fig.\,3. According to the London theory, the applied field $\textbf{H}_0$ induces a circumferential screening current persistently running along the film surface within a thin layer where the field is not zero; a transverse (perpendicular to $\textbf{H}_0$) cross-section of this current represents  a rectangle with an infinite length/width ratio ($\sim$ 5 mm/3 $\mu$m). 
But, how can electrons (supposed free) run along the straight path if they are constantly pushed sideways by the magnetic part of the Lorentz force? 
How can they make a U-turn at the ends keeping the constant speed, without canceling the law of inertia and following from it Newton's acceleration equation Eq.\,(2)? The London theory does not answer  these questions.

\begin{figure}
	\includegraphics[width=0.9\linewidth]{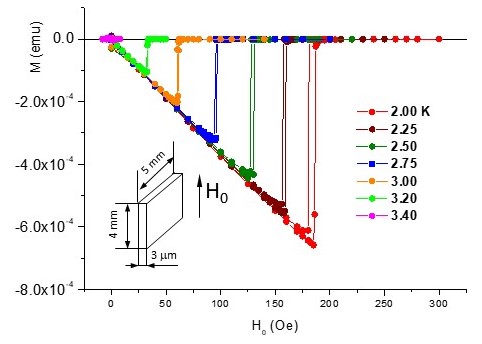}
	\caption{Magnetic moment of a 2.9-$\mu$m thick indium film in the parallel field $\textbf{H}_0$ \cite{IS-3}. The film  (deposited on a SiO$_2$/Si wafer) represents a rectangle schematically shown in the insert.} 
	\label{fig:epsart}
\end{figure}

Even clearer picture is as follows. Consider the cylindrical sample in Fig.\,1. Coming from the boundary conditions for $\textbf{B}$ \cite{Landafshitz_II,VK} and using the known formula for $\textbf{j}$, one can calculate average kinetic  energy $\epsilon_k$ of the field induced motion of one superconducting electron. This energy is $\epsilon_k=H_0^2/8\pi n_s$. From that one can find kinetic energy $E_k$ of the induced motion of all superconducting electrons in the penetration layer, which create the correct sample magnetic moment. This energy is
\begin{equation}
	E_k=\frac{H_0^2 V}{8\pi}\cdotp\frac{2\lambda_L}{\Re}, 
\end{equation}
where $\Re$ is the sample radius\footnote{This formula can be obtained without any calculations since in the theory superconducting electrons are active only in the penetration layer.  }. 

As mentioned, according to the energy conservation law, the magnetic energy of this sample $E_m(=H_0^2V/8\pi)$ equals the field induced kinetic energy of electrons $E_k$. However, in the London theory, as has just been calculated, $E_k\ll E_m$ which clearly contradicts the law\footnote{Note, that if each superconducting electron of the sample acquires kinetic energy $\epsilon_k$, then $E_k=n_s\epsilon_kV=VH_0/8\pi$: the law is met!}.  

The conflict of the London theory with the law of energy conservation was noted long ago by Shoenberg \cite{Shoenberg}. He paid attention to the fact that in a spherical sample in the MS the assumption of  circumferential current leads to an appearance of a Hall-like e.m.f. between points at the pole and the equator. If so, it would be possible to continuously draw energy  from the static magnetic field into a resistive circuit connecting these points, in contradiction with the energy conservation\footnote{Referring to Pippard, Shoenberg writes that to meet the law 
we must suppose 
that there is an opposite contact potential difference varying with the field in such a way to compensate this e.m.f. exactly \cite{Shoenberg}. An evident strangeness of such a supposition (the contact potential can be easily excluded by using the resistive circuit of the same metal as the sample, as is always done in studies of the Hall effect (see, e.g., \cite{Kikoin-31})) and other remarks throughout Shoenberg's book show that he and Pippard (see also a footnote on p.\,20 in \cite{VK}) well saw not only the merits of the London theory.}. 

After all, we mention a dilemma one more time demonstrating the inconsistency of the assumption of the circumferential screening current. Another dilemma stemming from this assumption is discussed in the Appendix.     

Consider a spherical sample in the MS. The external field $\textbf{H}_{ext}$ is parallel to the sample  surface  (as mentioned, it bends around the sample  in accord with the continuity of $B_n$) and its magnitude decreases down to zero with the polar angle $\vartheta$ as $H_{ext}=(3 H_0 \sin\vartheta)/2$, what was confirmed experimentally (see \cite{Shoenberg} for references). Therefore,  the London penetration depth $\lambda_L$ should decrease with $\vartheta$ as well because it must vanish at the pole where $H_{ext}=0$. However, in the London theory $\lambda_L$ does not depend on the field and therefore it is supposed staying the same regardless on changing $H_{ext}$ (see  footnote ($^{20}$)). This obvious contradiction stems from the assumption of the circumferential current.  
  
Any one of the listed inconsistencies (this list can be continued) is sufficient to  cast doubt on the London theory. However, still one more very strange thing is that the theory does not mention the necessary condition of the existence of persistent current: quantization of the angular momentum of its carriers. Below we will see how these issues can be resolved.

\section{MICRO-WHIRLS MODEL }

\section*{PROPERTIES OF THE MEISSNER STATE }

As known, diamagnetism of non-superconducting materials, being essentially a quantum phenomenon \cite{Tamm,Van Vleck}, is successfully described by the classical Langevin theory;  the volume magnetic susceptibility $\chi$ in this theory is identical to that in the quantum theory \cite{Feynman_Lectures}. 
A similar (but semi-classical) approach, as we will see below, may work for superconductors\footnote{In both cases this can be explained by the compensation of electron spins (either in molecules of conventional diamagnetics or in Cooper pairs of superconductors), which makes electrons responsible for the magnetic properties effectively spinless. The latter, in turn, excludes the appearance of Planck's constant in the formula for the magnetic susceptibility of these materials. }. An important advantage of the classical approach is transparency of the physical significance of the used concepts allowing to visualize processes underlying and determining properties under question. At the same time one should not forget that a comprehensive description of superconductivity is not possible without a full scale quantum-mechanical theory.

We start from the Bohr-Sommerfeld quantization condition, which ensures the dissipation-free electron motion over a closed path. At the same time,  we understand that in magnetic field the linear momentum and, hence, the angular momentum and the action, should be taken in the generalized form \cite{LL-field}. 

In superconductors electrons responsible for persistent current(s) are coupled in Cooper pairs \cite{Cooper}. Therefore, the quantization condition should be written for the paired electrons. Then the generalized action $\widetilde{S}_{cp}$ and the magnitude of  generalized angular momentum $\widetilde{\iota}_{cp}$ of one pair is

\begin{equation}
\widetilde{S}_{cp}=\oint \widetilde{\textbf{p}}_{cp}\cdot d\textbf{l}=2\pi R\widetilde{p}_{cp}=2\pi\widetilde{\iota}_{cp}=nh,
\end{equation}
where $\widetilde{\textbf{p}}_{cp}$ is generalized linear momentum of the pair, $R$ is radius of an orbital motion of paired electrons (the fact that the paired electrons are in the orbital motion will be seen slightly lower), $h$ is the Planck constant, and $n$ is a non-negative integer ($n=0, 1, 2, ...$).

In the ground state (i.e. at zero $T$ and $H_0$) $n$ takes the lowest  value $n=0$. Hence, the quantization condition (16) reads
\begin{equation*}
\oint \widetilde{\textbf{p}}_{cp}\cdot d\textbf{l}=0 \tag{16a}
\end{equation*}

Therefore (since in Eq.\,(16) $R\neq 0$ otherwise the sample magnetic moment would be always zero),
\begin{equation}
\widetilde{\textbf{p}}_{cp}=0. 
\end{equation}

This is the London rigidity principle written, however, for the paired electrons.

In zero field the generalized  linear momentum of the pair $(\widetilde{\textbf{p}}_{cp})_0$ equals its kinetic linear momentum $(\textbf{p}_{cp})_0$. Then from Eq.\,(17) it follows
\begin{equation}
(\widetilde{\textbf{p}}_{cp})_0=(\textbf{p}_{cp})_0=\textbf{p}_{10}+\textbf{p}_{20}=0,
\end{equation}
where $\textbf{p}_{10}(\equiv m\textbf{v}_{10})$ and  $\textbf{p}_{20}(\equiv m\textbf{v}_{20})$ are the kinetic linear momenta of each electron in the pair at zero field. 

This is  identical to the definition of Cooper pair \cite{Schrieffer} as a correlated state of two electrons with zero net kinetic linear momentum with respect to their center of mass and zero net intrinsic magnetic moment (spin). The latter circumstance explains the absence of the spin term in Eq.\,(16)\footnote{The zero spin of paired electrons follows from the requirement of thermodynamics, since the system of units with zero spin has lesser free energy. The same provides the thermodynamic justification of the profitability of the paired state of conduction electrons since the pairing makes possible to null the spin. }. 

On the other hand, Eq.\,(18) implies that in zero field the center of mass of the paired electrons is at rest (with respect to the sample) and electrons in each pair, separated by the coherent length $\xi$\footnote{At zero temperature $\xi$ corresponds to the Pippard/BCS coherence length $\xi_0$, which is close to the GL coherence length at this temperature \cite{Tinkham}. As we will see, $\xi$ does not depend on the field, but it does depend on temperature.}, orbit their center of mass, like, e.g., proton and electron in a hydrogen atom\footnote{Apparently, that $\textbf{p}_{10}$ and $\textbf{p}_{20}$ in Eq.\,(18) can not be directed toward each other or vice versa, because in such case Cooper pairs would not be stable.  }. Therefore each Cooper pair possesses the kinetic angular momentum $\vec{\iota}_0$ and the  magnetic moment $\vec{\mu}_0=\gamma\vec{\iota}_0$, where $\gamma$ is the gyromagnetic ratio. Hence, in a magnetic field the pairs should precess and below we will see that this is indeed so.

Due to symmetry, in zero field the total magnetic moment of all pairs (the magnetic moment of the sample) is zero, i.e. 
\begin{equation}
M_0=\sum\vec{\mu}_0=0,
\end{equation}
where summation is taken over all pairs. 

In view of uniformity of the bulk properties of the MS, Eq.\,(19) holds for the unit volume as well as for a physically infinitesimal volume element $dV$.  
The latter in superconductors should be defined as a volume, which size is much smaller than the size of the volume taken by the S phase and much larger than the spacial inhomogeneity of microscopic currents, i.e. $\xi$.  

The condition Eq.\,(19) is similar  to the definition of a diamagnetic atom, where each electron possesses a nonzero orbital magnetic moment, whereas the moment of the entire atom is zero. 
   
Now we turn the applied field $\textbf{H}_0$ on keeping the sample at constant temperature. Then the field intensity inside the sample  rises from zero to $\textbf{H}$ and  each of the paired electrons experiences the action of the Lorentz force\footnote{Naturally, the Lorentz force acts on "normal" (non-coupled) conduction  electrons as well resulting in a weak Landau diamagnetism \cite{Landau30,LL_Stats}.  
Action of the Lorentz force on the atomic electrons and nuclei results in the normal diamagnetic response.} $\textbf{F}$, which is \cite{LL-field}   
\begin{multline}
\textbf{F}\equiv\frac{d \textbf{p}}{d t}=-\frac{e}{c}\left(\frac{\partial \textbf{A}}{\partial t}\right)+\frac{e}{c}\textbf{v}\times \textbf{H}=\\-\frac{e}{c}\left(\frac{\partial \textbf{A}}{\partial t}\right)+\frac{e}{c}\textbf{v}\times (\nabla \times \textbf{A}), 
\end{multline}
where $t$ is time, $\textbf{A}$ is the vector potential  of the magnetic field $\textbf{H}(=\nabla \times \textbf{A})$ acting on electron, and $\textbf{v}$ is the electron velocity, which magnitude is slightly less than the Fermi velocity $v_F$ due to condensation\footnote{The maximum difference between $v_F$ and $v$ is $\Delta v_{cond}=v_F-v_0\approx v_F(E_g/E_F)^{0.5}\lesssim 10^{-2}v_F$, where $v_0$ is velocity of the condensed (superconducting) electrons at zero field and temperature, and $E_g$ and $E_F$ are the energy gap at $T=0$ and the Fermi energy, respectively;  the numbers are taken from \cite{Kittel}.}. 

In Eq.\,(20)  Coulomb's term $-e\nabla\varphi$, where $\varphi$ is the electrostatic potential, is omitted in view of the absence of the applied 
electrostatic field\footnote{In the London theory the applied electrostatic field penetrates the superconductor following the same law as that for the magnetic field. An experimental attempt undertaken by H. London to reveal this effect yielded zero result \cite{HLondon-36}.}.

Here we need to make a brief stop. As known (see, e.g.,  \cite{VK,Jackson}), the field  $\textbf{H}$ is a potential-like field. It can be described using the magnetic scalar potential $\Psi$ and ``magnetic charges" resided on the sample boundary; a surface density of these charges equals the normal component of magnetization $\textbf{I}_n$. Respectively, if one considers a volume element $dV$ which includes a piece of the boundary with $\textbf{I}_n\neq 0$, the total flux of the field $\textbf{H}$ through its surface is not zero. This implies that $\nabla\cdot \textbf{H}\neq 0$ and hence the concept of the vector potential is inapplicable to the field $\textbf{H}$ \textit{in this region}. However, if $dV$ is totally inside the sample (it can touch but not cross its boundary), than $\textbf{H}$ becomes divergence-less field and therefore it can be described by a vector potential $\textbf{A}$ defined through the relationship  $\nabla\times \textbf{A} = \textbf{H}$. Note that, unlike $\textbf{H}$, the induction $\textbf{B}$ is  a divergence-less field everywhere. 

One more thing: since inside real bodies, including samples in our model, $\textbf{B}$ and $\textbf{H}$ are different, one has to distinguish the vector potential for $\textbf{H}$ and for $\textbf{B}$. So in our notations $\textbf{\textbf{A}}$ is the vector potential of the field $\textbf{H}$ inside the sample. Below we will see how $\textbf{A}$ is related to the vector potential $\textbf{A}_B$ defined by the relationship $\textbf{B}=\nabla\times \textbf{A}_B$. As usual, we presume the  supplementary condition Eq.\,(10)  for $\textbf{A}$ and $\textbf{A}_B$. 

The first term in  the right-hand side of Eq.\,(20), referred to as the electric force $\textbf{F}_E$, is a component of the Lorentz force  due to the vortex electric field defined as
\begin{equation}
\textbf{E}\equiv \frac{\textbf{F}_E}{e} =-\frac{1}{c}\left(\frac{\partial \textbf{A}}{\partial t}\right).
\end{equation}

The force $\textbf{F}_E$ (existing  while the magnetic field is changing) does the work resulting in the change of kinetic energy of electrons and  in the appearance  of the induced magnetic moment in each pair. This is pretty much the same as the familiar phenomenon of magnetization in regular diamagnetics (see, e.g., \cite{Purcell,Griffiths,Feynman_Lectures,Tamm}). 

When the vector potential changes from zero to $\textbf{A}$, which corresponds to the change of the field intensity from zero\footnote{We choose $A=0$ at $H=0$.} to $\textbf{H}$, the force $\textbf{F}_E$ changes velocity of electrons in the pair from $\textbf{v}_0$ to $\textbf{v}'$. The difference $\textbf{v}_i= \textbf{v}'-\textbf{v}_0$ (the velocity induced due to the applied magnetic field), according to Eq.\,(21),  is\footnote{Note that $\textbf{v}_i$ in this formula is identical to $\textbf{v}$ in the London theory (Eq.\,(12)). This is the reason of  the equality of the sample magnetic moment in our model (calculated below) and that in the London theory. }
\begin{equation}
\textbf{v}_i=-e\frac{\textbf{A}}{cm}. 
\end{equation}

There are two things in Eq.\,(22) to be pointed out: (i) like in regular diamagnetics \cite{Purcell}, the time dropped out, which means that the induced velocity $\textbf{v}_i$ is the same regardless on the rate of the field change; (ii) $\textbf{v}_i$ is  parallel to $\textbf{A}$, implying that $\textbf{v}_i$ is tangential to the line of  vector potential laying by definition in the plane  perpendicular to $\textbf{H}$.  Since $e<0$, the latter means that the magnetic moment induced in each Cooper pair is \textit{always} diamagnetic, in accord with requirements of electrodynamics \cite{Tamm} and thermodynamics \cite{VK}.  

The second term in the right hand side of Eq.\,(20), referred to as the magnetic  force $\textbf{F}_M$,  is a component of the Lorentz force perpendicular to the electron velocity. Hence, $\textbf{F}_M$ does no work and therefore it does not affect the electron kinetic energy and the magnitude of its magnetic moment. Then, what does $\textbf{F}_M$ do? 

To answer this question, we note  that, (i) since the uniform magnetic field cannot change position of the pair's center of mass, $\textbf{F}_{M1}=-\textbf{F}_{M2}$ (those are the magnetic forces acting on the 1st and 2nd electron in the pair),  and (ii) $\textbf{F}_{M1}$ and $\textbf{F}_{M2}$ are non-central forces. Hence, they make a couple resulting, due to non-zero $\vec{\iota}_0$, in precession of $\vec{\mu}_0$ relative to the vector $\textbf{H}$, as schematically shown in Fig.\,4. 

\begin{figure}
	\includegraphics[width=0.8\linewidth]{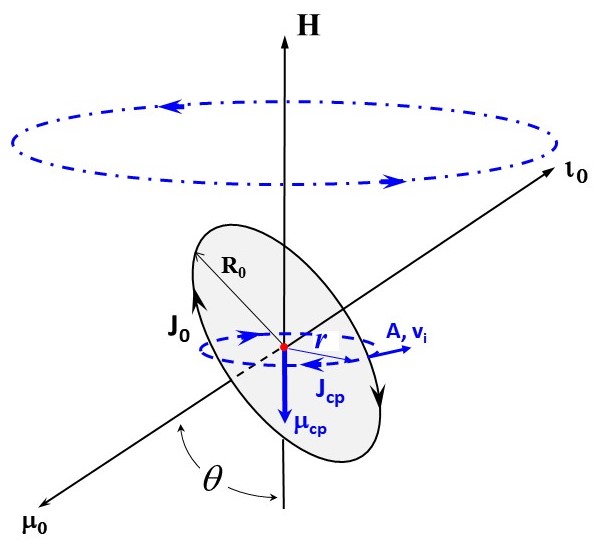}
	\caption{Schematics of the Cooper pair precession in the magnetic field $\textbf{H}$. Vectors $\vec{\iota}_0$ and $\vec{\mu}_0$ are, respectively, the angular and magnetic moments of the pair caused by the the circular current $J_0$ formed by the coupled electrons in zero field; this current is caused by the orbital motion of the paired electrons about their center of mass (a red dot); $R_0$ is the orbit radius. A dot-dashed circle designates the path of the tip of the precessing $\vec{\iota}_0$. A dashed circle depicts the  field induced current $J_{cp}$; it is also a line of the vector potential directed opposite to the current $J_{cp}$. $\textbf{A}$ and $\textbf{v}_i$ designate the vector potential and the induced velocity of one electron, respectively; $r$ is the radius of the induced current; and $\vec{\mu}_{cp}$ is the induced magnetic moment of the pair. Depicted by a single arrow vectors $\textbf{A}$ and $\textbf{v}_i$ are proportional but not equal to each other.	} 
	\label{fig:epsart}
\end{figure}

The angular velocity of precession, referred to as Larmor frequency, is 
\begin{equation}
\textbf{o}=-\gamma \textbf{H}=-\frac{eg_L}{2mc}\textbf{H},
\end{equation}
where $\gamma$ is the gyromagnetic ratio and $g_L$ is the Lande factor.

In superconductors  $\gamma$ was  measured in the  experimental masterpiece of I.  Kikoin\footnote{Academician Isaak Konstantinovich Kikoin was a brilliant physicist, the deputy of Kurchatov in the Soviet atomic project, founder and director of the Division of molecular physics of the Kurchatov institute, in which the author was working for many years.} 
and Goobar \cite{Isaak-1}; detailed report was published in \cite{Isaak-2}. The measurements were performed on a high purity ZFC lead spherical samples of 3-4 mm in a diameter. To ensure the absence of the frozen flux, the earth field was compensated down to $\lesssim 5\cdot 10^{-4}$ Oe. The measured Lande factor is $1 \pm 0.03$.  I. K. concluded that  (1) ``the magnetization of superconductors, in any case, is caused not by electron spin, but by closed electron currents" and  (2) this "can be due to microscopic closed currents, although their origin \tg{so far} is not known" \cite{Isaak-2}.

As follows from the Larmor theorem, if magnitude of the induced electron velocity $v_i$ is much less than $v_0$, precession of the electron orbit is equivalent to undisturbed orbital motion in the field absence (i.e. with fixed $\vec{\mu}_0$) plus an additional (field induced) circular motion with the angular velocity $\textbf{o}$ and radius $r$ proportional to $R_0$, which leads to appearance of the diamagnetic moment $\vec{\mu}_i$ \cite{Tamm,Feynman_Lectures}. The same result can be obtained without direct involvement of the Larmor theorem \cite{Purcell,Landsberg,Griffiths}. The latter approach explicitly shows that the diamagnetism results from the changing magnetic field, in full accordance with the Faraday law. On the other hand, the invariability of $\vec{\mu}_0$ means that condition Eq.\,(19) holds both in the absent and presence of the magnetic field\footnote{This can be also viewed as follows. In the uniform magnetic field all pairs precess synchronously because precession is a motion (the only one of a kind) occurring without inertia. Then, since $\vec{\mu}_0$ stays unchanged, orientations of the pairs' magnetic momenta with respect to each other stay unchanged either, meaning that $\sum\vec{\mu}_0$ remains zero.}.
\begin{figure}
	\includegraphics[width=0.7\linewidth]{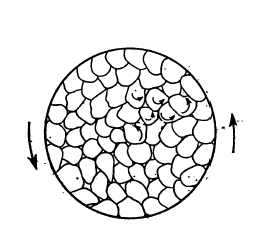}
	\caption{A cross sectional view of a conventional diamagnetic sample showing induced bound currents caused by precession of  the atomic electron orbits; the field is directed into the page (copied from \cite{Tamm}). This picture is  identical to the induced currents caused by  precession of Cooper pairs in a superconducting sample in the Meissner state.} 
	\label{fig:epsart}
\end{figure}

Thus, we arrive to conclusion that the net effect of the  magnetic field is the induced circular motion of the paired electrons in the plane perpendicular to $\textbf{H}$. On the other hand, since changing $|\textbf{H}|$ changes  only the magnitude of $\textbf{v}_i$, the \textit{radius of the induced motion  $r$ does not depend of the field}. 

This is very close to the picture of induced bound currents  in regular diamagnetics schematically shown in Fig.\,5. In our case, like in \tg{the normal diamagnetics},  the induced currents mutually compensate  each other in the sample bulk, leaving an uncompensated surface current caused by electrons bound in Cooper pairs. Then the magnetic moment of the sample is exactly the same as the moment produced by a continuous (circumferential) surface current \cite{Tamm,VK,Feynman_Lectures,Griffiths,Purcell}.

Now let us check what happens to $\widetilde{\textbf{p}}_{cp}$ in the field. For that we go back to the quantization condition (16a) and look what is going on  when the field $\textbf{H}_0$ is turned on, i.e., it changes from zero to $\textbf{H}_0$ over some time interval.  
Then,  inside the sample the field intensity changes from zero to $\textbf{H}$ and, correspondingly, the vector potential  changes from zero to $\textbf{A}$ over the same time (more correctly to say that in the reversed order). From Eq.\,(20) we find that the kinetic linear momentum changes for $-e\textbf{A}/c+\int\textbf{F}_{M}dt$ and therefore the generalized linear momentum of the single Cooper pair in the field is 
\begin{multline}
\widetilde{\textbf{p}}_{cp}=\widetilde{\textbf{p}}_1+\widetilde{\textbf{p}}_2=\left[\left(\textbf{p}_{10}-\frac{e}{c}\textbf{A}_1 + \int\textbf{F}_{M1}dt\right)+\frac{e}{c}\textbf{A}_1\right]+\\ \left[\left(\textbf{p}_{20}-\frac{e}{c}\textbf{A}_2+\int\textbf{F}_{M2}dt\right)+\frac{e}{c}\textbf{A}_2\right]=\textbf{p}_{10}+\textbf{p}_{20}=0,
\end{multline}
where $\textbf{A}_1$ and $\textbf{A}_2$ are the vector potentials experienced by the first and second electron in the pair, respectively. 
   
Thus we see that, as predicted by F. London for superconducting electrons,  $\widetilde{\textbf{p}}_{cp}=0$ regardless  on the presence or absence of the magnetic field. Q.E.D.   

Now let us calculate magnetic proprieties of samples in the  MS, i.e. of the ellipsoidal bodies in which $\widetilde{\textbf{p}}_{cp}$ of the paired electrons  is zero.   
For that we have to choose an appropriate form of \textbf{A} and to link it with $\textbf{H}_0$.

A uniform field  $\textbf{H}=H\hat{\textbf{z}}$ ($\hat{\textbf{z}}$ is unit vector along the $z$-axis) can be described by  $\textbf{A}$ of the following forms or gauges (see, e.g., \cite{Kroemer})
 
\begin{equation}
\textbf{A}=-Hy\hat{\textbf{x}},
\end{equation}
\begin{equation}
\textbf{A}=Hx\hat{\textbf{y}},
\end{equation}
and 
\begin{equation}
\textbf{A}=\frac{1}{2}\left( -Hy\hat{\textbf{x}}+Hx\hat{\textbf{y}}\right) =\frac{1}{2}\textbf{H}\times \textbf{r},
\end{equation}
where $\hat{\textbf{x}}$ and $\hat{\textbf{y}}$ are unit vectors in the $x$- and $y$-direction, respectively; and $\textbf{\textbf{r}}$ is the radius vector lying in the  $xy$ plane perpendicular to $\textbf{H}$ with an origin in an arbitrary point of this plane.

\begin{figure}
	\includegraphics[width=0.7\linewidth]{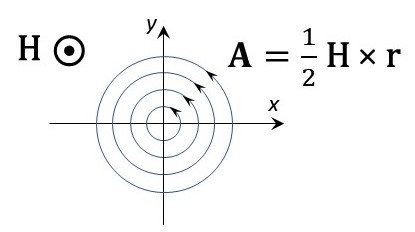}
	\caption{Lines of the vector potentials  $\textbf{A}$ in the circular gauge for a uniform magnetic field $\textbf{H}$  (Eq.\,(27)). $\textbf{H}$ is directed along the $z$-axis (toward the reader). As always,  the line of $\textbf{A}$ is the directional line tangential to  the vector $\textbf{A}$ in each of its point.} 
	\label{fig:epsart}
\end{figure}

The vector potentials Eqs.\,(25)-(27) are equivalent\footnote{They differ from each other by a gradient of a function of coordinates. For example, $\textbf{A}$ in \tg{Eq.\,}(27) differs from $\textbf{A}$ in \tg{Eq.\,}(25) by $\nabla (xyH/2)$.} in a sense  that they represent the same field $\textbf{H}$. However, as seen from Eq.\,(22), in superconductors $\textbf{A}$ of different gauges lead to different $\textbf{v}_i$ and, therefore, to different magnetic moment of the sample. 
This implies that the vector potential in superconductors is not gauge-invariant, as it also takes place in the London and BCS theories\footnote{F. London admits this fact noting that due to this reason Eq.\,(\tg{9}) cannot be generally valid \cite{London50}. In the BCS theory lack of the gauge invariance is attributed to the approximate character of the theory \cite{Schrieffer}. At the end of his book \cite{London50} F. London shows that in \tr{a} quantum theory the gauge invariance can be preserved if the gauge transformation is accompanied by a corresponding transformation of the wave function of superconducting electrons.} \cite{Schrieffer}. This is an additional confirmation of the fact that the vector potential is not just a mathematical fiction but a real and primary characteristics of the magnetic field, as demonstrated by the  Aharonov-Bohm effect \cite{Aharonov-Bohm,Feynman_Lectures}.  

In our case the choice of $\textbf{A}$ is quite obvious: due to uniformity of $\textbf{H}$, all directions in planes perpendicular to $\textbf{H}$ are equivalent, so an appropriate $\textbf{A}$ is that in Eq.\,(27), which is referred to as the vector potential of the circular gauge. This also follows from the fact that the induced currents and lines of the vector potential $\textbf{A}$ must make closed loops. The later is consistent with the condition $\nabla\cdot\textbf{A}=0$. The lines of the vector potential of the circular gauge are shown in Fig.\,6. 

After turning on the applied field $\textbf{H}_0$, the field intensity and the vector potential inside the sample after a short relaxation time become $\textbf{H}$ and $\textbf{A}$, respectively. Then, using Eq.\,(22), we write 
\begin{equation}
m\textbf{v}_i=-\frac{e}{c}\textbf{A}=-\frac{e}{2c}\textbf{H}\times \textbf{r},
\end{equation}

In the scalar form, taking into account that $e<0$, Eq.\,(28) reads
\begin{equation}
mv_i=\frac{e}{2c}Hr,
\end{equation}  
where $e$ is the absolute value of electron charge.

Hence, magnitude of the angular velocity of the induced circular motion is
\begin{equation}
\omega=\frac{v_i}{r}=\frac{e}{2mc}H.
\end{equation} 

We see that the angular velocity is equal to the classical Larmor frequency ($g_L=1$).  This confirms that we really deal with the precession of paired electrons with zero total spin, since non-paired electrons (as any other charges) in a magnetic field circulate with a so called cyclotron frequency $\omega_c=(e/mc)H$ \cite{Purcell,LL_Stats,Landau30}.

The induced current per one electron in the pair $J_i$ is
\begin{equation}
J_i=\frac{e}{2\pi}\omega={\frac{e^2}{4\pi mc}}H.
\end{equation}

From Eqs.\,(30) and (31) we see that neither induced angular velocity $\omega$, nor the induced current $J_i$ depends on $r$. However, it is not the case for the induced magnetic moment and the corresponding change of kinetic energy of the paired electrons: they both depend on $r^2$. On the other hand, $r$ for Cooper pairs with different orientation of $\vec{\mu}_0$ should be different. So what we want to know is the mean square $\langle r^2\rangle$,  which we will denote as $r_i^2$.

In conventional diamagnetics $r_i$ is calculated from the Langevin formula for the magnetic susceptibility $\chi$ and corresponding experimental data \cite{Van Vleck}. Calculated in this way values of $r_i$ are shown in Table I. We see that in many substances the values of $r_i$ are quite close,   between 1.5-2.0 \AA, however in some materials, e.g., copper, it is less than 1 \AA, whereas in bismuth and graphite it is significantly greater. In superconductors $r_i$ can be found  as follows.

\begin{table}
\caption{The root mean square radius $r_i$ of the induced current in conventional diamagnetics calculated from Langevin's formula for  $\chi$ \cite{Tamm}. The values of $\chi$ are taken from \cite{Grigoriev}; $\chi$ for pyrolytic graphite is taken from Wikipedia. For metals (Cu, Bi and graphite) it is assumed that one electron of each atom is unbound (is in the conduction zone).
}
\vspace{1mm}
\begin{tabular}{c c c c}

\hline\\[0.05ex]

	Substance       &Formula \hspace{3mm} & \hspace{4mm}   $10^6\chi$ \hspace{5mm}  &$r_i$, \AA \\ [1ex]

\hline\\
	Copper          & Cu     & -0.771           & 0.68 \\[0.5ex]
    Sodium Chloride & NaCl   & -1.121           & 1.60 \\[0.5ex]
    Sulfur          & S      & -0.956           & 1.52 \\[0.5ex]
    Diamond         & C      & -1.543           & 1.52 \\[0.5ex]
    Graphite        & C      & -10.81           & 4.13 \\[0.5ex]
    Pyrol. graphite & C      & -31.8            & 9.51 \\[0.5ex]
    Nitrogen (liq)  & N$_2$  & -0.410           & 1.55 \\[0.5ex]
    Bismuth         & Bi     & -19.951          & 3.53 \\[0.5ex]
    Water           & H$_2$O & -0.720           & 1.96

\end{tabular}

\end{table}

An average induced magnetic moment per one electron in Cooper pairs is
\begin{multline}
\mu_i=\frac{1}{c}J_i\pi r_i^2=\frac{e^2r_i^2}{4mc^2}H=\frac{e^2n_s4\pi}{mc^2}\left(\frac{r_i}{2}\right)^2\frac{H}{4\pi n_s}=\\\frac{(r_i/2)^2}{\lambda_L^2}\frac{H}{4\pi n_s}.
\end{multline}

For simplicity, let us consider a sample of cylindrical geometry ($\eta=0$). For this geometry (i) the demagnetizing field $\textbf{H}_d(=4\pi\eta\textbf{I}$) is zero and therefore $\textbf{H}=\textbf{H}_0-\textbf{H}_d=\textbf{H}_0$\footnote{As mentioned above, this also follows from the boundary condition for the tangential component of the field $\textbf{H}$ and from the Poison theorem \cite{VK}.}; (ii) the outside magnetic field produced by the magnetized sample is absent and therefore, as was already mentioned, the sample magnetic energy $E_m$ equals the field induced change of kinetic energy of the paired electrons $\Delta T$ \cite{VK}.

Since magnetic moments induced in all Cooper pairs are parallel,  the  magnetic moment of our sample is
\begin{multline}
M=n_{cp}V\mu_{cp}=n_sV\mu_i=\left(\frac{(r_i/2)}{\lambda_L}\right)^2\frac{V}{4\pi}H=\\ \left(\frac{(r_i/2)}{\lambda_L}\right)^2\frac{V}{4\pi}H_0,
\end{multline}
where $n_{cp}( =n_s/2)$ is number density of the pairs.

Referring Eq.\,(13), we know  that magnitude of  the magnetic moment of the cylindrical sample in the MS is   
\begin{equation*}
M=\frac{V}{4\pi}H_0.
\end{equation*}

Hence we see, that our model meets thermodynamics and the experiment provided the radius $r_i(\equiv\sqrt{\langle r^2\rangle})$  of the induced circular motion of the coupled electrons is
\begin{equation}
r_i=2\lambda_L.
\end{equation}

Since $r$ does not depend on the field, $r_i$ does not depend on the field either. Hence, as was  assumed in the London theory, $\lambda_L$ in our model does not depend on the field at constant temperature. 

Next, one can show that like in regular diamagnetics \cite{VK} the change of kinetic energy  of  the superconducting electrons $\Delta T$ is the sum of the field induced kinetic energies of each such electron\footnote{This follows from the validity of Eq.\,(19) at $\textbf{H}_0\neq0$ and the uniformity of the field $\textbf{H}$ within the sample.} $\epsilon_i=mv_i^2/2$, i.e.
\begin{equation}
\Delta T=E_k=\epsilon_i n_s V=\epsilon_{cp} n_{cp} V,
\end{equation}
where  $\epsilon_{cp}=2\epsilon_i$ is the average kinetic energy of the induced motion of electrons in one pair.

Using Eq.\,(29), we write
\begin{multline}
\epsilon_i=\frac{mv_i^2}{2}=\frac{(mv _i)^2}{2m}=\dfrac{e^2H^2r_i^2}{8c^2m}=\\\left(\dfrac{4\pi e^2 n_s}{mc^2}\right)\left(\frac{r_i}{2}\right)^2\frac{H^2}{8\pi n_s}=\left(\frac{(r_i/2)}{\lambda_L}\right)^2\frac{H^2}{8\pi n_s}.
\end{multline}

Thus, the  change of kinetic energy of the paired electrons is
\begin{equation}
\Delta T = \epsilon_i n_sV=\left(\frac{(r_i/2)}{\lambda_L}\right)^2\frac{H^2}{8\pi}V.
\end{equation}  

In our sample $H=H_0$ and, according to the energy conservation, $\Delta T=E_m=H_0^2V/8\pi$ (see Eq.\,(14)). This confirms that $r_i=2\lambda_L$. 

Next, we calculate the gyromagnetic ratio coming from its definition. Using Eqs.\,(29), (32) \tg{and (34)} we write 
\begin{equation}
\gamma\equiv\frac{M_i}{L_i}=\frac{\mu_i n_sV}{\iota_i n_s V}=
\frac{\mu_i}{mv_ir_i}=\frac{e^2r_i^2H}{4mc^2}\frac{2c}{er_i^2H}=\frac{e}{2mc},
\end{equation}
where $M_i$ and $L_i$ are magnitudes of the induced magnetic  and angular momenta of the sample, respectively; and $\iota_i$ is an average field-induced angular momentum per one electron. 

We see that $\gamma$ is fully consistent with the experimental result of I. Kikoin and Goobar \cite{Isaak-2,Isaak-1}. Q.E.D. 

After all, let us calculate the induction in the sample interior. Using Eqs.\,(32) and (\tg{34}), and the fact that $\vec{\mu}_i$ is negative, we obtain
\begin{equation}
\textbf{B}=\textbf{H}+4\pi \textbf{I}=\textbf{H}+4\pi(\vec{\mu_i}n_s)=\textbf{H}-4\pi \frac{\textbf{H}}{4\pi n_s}n_s=0.
\end{equation}
Q.E.D.

Correspondingly, the magnetic permittivity $\mu_m\equiv B/H$ and susceptibility per unit volume  $\chi$ of the S phase are zero and $-1/4\pi$, respectively, as it should \cite{Meissner,Shubnikov}.  

Naturally, due to microscopic character of the induced currents, there is no problem in establishing the Meissner condition ($B=0$) in samples/domains of any shape, as soon as the field $\textbf{H}$  is uniform. The latter is indeed so in ellipsoidal samples \cite{VK}, regardless whether they are in the MS or in the inhomogeneous equilibrium states, i.e., in the mixed state of type-II and in the intermediate state of type-I superconductors. 

\begin{figure}
\includegraphics[width=0.7\linewidth]{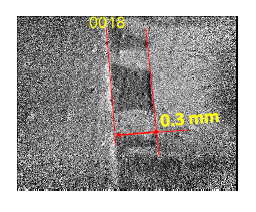}
\caption{ A photonmagnetic image of superconducting domains (dark areas) in a 2.5-$\mu$m thick indium film sample in the intermediate state at temperature 3 K in the tilted field.  The sample (outlined by the red lines) represents a stripe of 0.3 mm in width and 1 mm in length. An out-of-plane component of the applied field $H_{0\perp}=$ 0.5 Oe and the in-plane component  $H_{0\parallel}=$ 50 Oe. As one can see the central S domain has a huge ratio of the lateral size to the thickness ($\sim$0.3 mm/2.5 $\mu$m) and its shape has nothing in common with the ellipsoid of revolution. Other images obtained with this sample are available in \cite{IS-1}.} 	\label{fig:epsart}
\end{figure}

This explains why the Meissner state  is observed only in the ellipsoidal bodies 
and, on the other hand, a vast  variety of (never spheroidal!) shapes of the S-domains in the intermediate state \cite{Huebener}, one example of which is shown in Fig.\,7. Note that the specific shape of domains in  both in-plane and out-of-plane cross sections of a pinning-free sample is  dictated by the thermodynamic profitability (i.e. by the minimal free energy) for the entire sample which may include quite a large space (as compared to the sample volume) adjacent to it \cite{IS-3}. 

The above consideration applies to the samples cooled in zero field, whereas  the Meissner effect is about the field-cooled (FC) samples. So, what happens with the FC samples?

Upon lowering temperature below $T_c(H_0)$ in the fixed field $H_0< H_{c1}(1-\eta)$, a temperature dependent fraction of conduction electrons condenses forming stable Cooper pairs. This means that speed of these electrons drops from $v_F$ to $v_0$, each pair starts orbiting its center of mass and, being in the field, the pairs precess\footnote{Recall that precession is the motion occurring without inertia.}. 
Like in regular diamagnetics, the latter leads to establishing magnetization $\textbf{I}$, the field intensity $\textbf{H}=\textbf{H}_0-4\pi\eta \textbf{I}$ and the induction $\textbf{B}=\langle \textbf{h} \rangle=\textbf{H}+4\pi\textbf{I}$, where $\langle \textbf{h} \rangle$ is the average microscopic field. Hence, after the short relaxation time needed to establish the field $\textbf{H}$, the environment inside the FC sample becomes the same as that in the ZFC sample. Thus, this model meets the Meissner effect indeed\footnote{In the London theory the Meissner effect is achieved by postulating $B_\infty=0$ (Eq.\,(6)).}. 

One more remark. As mentioned above, radius of the electron orbit $R_0$ in precessing Cooper pairs  is fixed (i.e. it does not depend on the field) or the Larmor theorem is exact if $v_i\ll v_0$  \cite{Purcell,Tamm}. Taking typical value of $H_{c1}\sim 100$ Oe and $\lambda_L \sim 10^{-6}$ cm, from Eq.\,(25) one finds $v_i\sim 10^2$ cm/s, which is six orders of magnitude less than $v_0\approx v_F\sim 10^8$ cm/s \cite{Kittel}. So, there is no doubt  that $R_0$ is the field independent quantity at constant temperature.

Now, when we worked out with the current induced in a  single Cooper pair, let us try to reconstruct the current structure of the Meissner state. We understand that all induced currents form  identical circular loops with the rms radius $r_i$  laying in parallel planes perpendicular to the field $\textbf{H}$. How these currents are arranged with respect to each other\footnote{We remind that Cooper pairs strongly overlap \cite{Schrieffer}, which, nevertheless, makes no effect on either stability or mobility of each pair. Recalling the quantum-mechanical nature of electrons, this is similar to the fact that overlapping myriads of electromagnetic waves around us do not prevent us from clearly seeing different objects and enjoying music broadcast by various radio stations.}?    

Coming from symmetry, one can expect two options: either complete chaos or complete order. Thermodynamics suggests that the second option is more preferable since the system of the ordered currents has lesser free energy. This is consistent with experimental facts that entropy of the sample in the S state is less 
than that of the N state (see \cite{Shoenberg} for references), and that \tg{Abrikosov} vortices in the mixed state form an ordered 2D structure of the maximal symmetry (hexagonal lattice)\footnote{Below we will see that Abrikosov vortices are holes in the network of the ordered induced currents of the Meissner phase.} \cite{Essmann}.

So, the induced currents of the MS can be  modeled as an ordered 2D structure of cylindrical micro-whirls (similar to the quantized vortices in superfluid helium \cite{Feynman57}) resembling densely packed and (as we will see next) very tightly ``wound” micro-solenoids aligned with $\textbf{H}$.   Length of each whirl/solenoid equals  the sample size along direction of $\textbf{H}$ and its rms diameter is $2r_i=4\lambda_L$. Since the solenoids are parallel to each other, they do not interact.  
This is consistent with the fact that the internal energy of a \tg{sample} in the MS is just the sum of kinetic energies $\epsilon_i$ (Eq.\,(37)) and it does not contain the term(s) responsible for interaction \cite{VK}. This is also consistent with experimental data evidencing that the Abrikosov vortices do not interact with each other \cite{MS}. After all, this is consistent with the  NMR experimental data (see, e.g., \cite{Knight,Reif,Ishida}) showing that the Knight shift in superconductors when extrapolated to $T=0$ is not zero, in contrast to what is expected in the BCS theory  \cite{BCS,Schrieffer}.  

Note that the picture of ordered currents corresponds to the long-range ordering of superconducting electrons, which follows from the postulate of zero entropy of the two-fluid model of Gorter and Casimir and was expected by F. London, based on his discovery of the rigidity principle. On the other hand, the fact that  $r_i^2$ is the average quantity (over possible angles $\theta$) means that electrons in precessing Cooper pairs experience action of the vector potential averaged over a space with dimension on the order of the pairs' size $\xi=2R_0$. This corresponds to the Pippard/BCS non-locality principle \cite{Pippard53,Schrieffer}. 

Now let us ask what is the spacing between the solenoid's  turns or between the induced current loops in direction of $\textbf{H}$? One can estimate it as follows.   

Consider the cylindrical sample in the  Meissner state in the field $H_0$ (like, e.g.,  one shown in Fig.\,1).  The magnitude of the linear density of the surface current\footnote{Unlike the surface current in the London theory, in our case this current is formed by electrons bound in stationary Cooper pairs, so there are no either electrons or pairs running along the surface.} $g$, calculated from the boundary condition for the tangential component of the induction $B_t$ (see, e.g., \cite{VK}), is 
\begin{equation}
g\equiv \frac{J}{L}=\frac{cH_0}{4\pi}=cI,
\end{equation}
where $J$ is the surface current and $L$ is the length of our cylinder. 

Since $I=\mu_i n_s$, we write
\begin{equation}
g=c\mu_in_s=c\frac{J_i \pi r_i^2}{c}n_s=J_{cp}\frac{\pi r_i^2}{A_cL}(n_{cp}V),
\end{equation}
where $J_{cp}=2J_i$ is the induced current per one Cooper pair, and $A_c$ and $V$ are the cross-sectional area and the volume of our sample, respectively.

Let us denote the number of loops in the sample cross  sectional area $A_c$ (perpendicular to $\textbf{H}$) as $N_\perp=A_c/\pi r_i^2$ and the number of loops along $L$ (parallel to $\textbf{H}$) as $N_\parallel$. The total number of the loops is equal to the number of pairs $N_{cp}$, which is $N_{cp}=n_{cp}V=N_\perp N_\parallel$. Then, 
\begin{equation}
g=J_{cp}\left(\frac{N_{cp}}{N_\perp}\right)\frac{1}{L}=2J_i\frac{N_\parallel}{L} 
\end{equation}

The last fraction is the number of the loops per unit length. Denoting $N_\parallel/L=n_\parallel$ and using Eq.\,(31) we find that the distance between the loops along direction of $\textbf{H}$ or the spacing between the induced current loops $\Delta$ is  
\begin{multline}
\Delta=\frac{1}{n_\parallel}=2J_i\frac{4\pi}{cH_0}=\dfrac{2e^2H_0}{4\pi mc}\frac{4\pi}{cH_0}=\frac{2e^2}{mc^2}=\\5.6\cdot 10^{-13} cm \approx 6\,fm.
\end{multline}

So, the loops are very tightly packed and $\Delta$ is a universal number, about three times the size of a proton (1.7 fm). In terms of $\lambda_L$ the spacing is 
\begin{equation}
\Delta=\frac{1}{\lambda_L^2 n_s 2\pi}=\frac{1}{4\pi n_{cp}\lambda_L^2}.
\end{equation} 

Next, what is the penetration depth $\lambda$ (the width of the surface layer with $B\neq0$) in this model: is it $\lambda_L$, $2\lambda_L$ or something else? It should be a combination of $r_i$ and $R_0$  each of which is proportional to $\lambda_L$, but this question is so far open. However, in any case one can  state that $\lambda$ is proportional to $\lambda_L$. On that reason, since $\lambda_L$ does not depend on the field, $\lambda$ does not depend on the field either, in accord with results of the microwave measurements of Pippard\footnote{In \cite{Pippard50} Pippard reported results of measurements $\Delta\lambda_L /\lambda_L$, the relative variation of \tg{the effective penetration depth} $\lambda_L$ vs applied field changing from zero to $H_c$ \tg{at fixed} temperatures. The experiment was conducted using microwave resonator with wavelength 3 cm (10 GHz). A small and non-monotonic temperature dependent increase (between about 0.002 and 0.03) of $\Delta\lambda_L /\lambda_L$ was found. Pippard noted that due to assumptions made the correct variation of \tg{$\lambda_L$} is probably smaller, so he concluded that $\lambda_L$ can be considered as being independent of the field. Later it was demonstrated that the high-frequency radiation in the Pippard's resonator  disturbs equilibrium distribution of the current carriers near the sample surface thus leading to an additional error not accounted by Pippard. More about this experiment will be said in Appendix.} \cite{Pippard50} and our LE-$\mu$SR data \cite{muons-18} (see Sec.\,VI).

All what was discussed so far is related to the samples at $T=0$. But if $T\neq0$, how will this affect the considered properties? The short answer is nohow. 

Indeed, like in regular diamagnetics (see problems 2.2 and 2.3 in \cite{VK}), the entropy $S_{cp}$ of the S-fraction of the conduction electrons, i.e. of the ensemble of Cooper pairs,  in our model (below we will call it \tg{the} micro-whirls (MW) model) is zero due to condition Eq.\,(19) and complete ordering of the field induced magnetic moments of Cooper pairs.  But  according to the Third law (Nernst's theorem), temperature of a statistical ensemble with zero entropy is zero. Therefore, the temperature of the ensemble of Cooper pairs $T_{cp}$ is zero regardless on the sample temperature $T$.  Hence, all results obtained in this section hold in  the whole temperature range of the existence of Cooper pairs.  Respectively, all calculations and formulae of this section hold at $0\leqslant T<T_c$.  In other words, the paired electrons are in the ground state in the entire temperature and field range of the S phase existence.     
 
Referring back to the zero-entropy postulate of Gorter and Casimir, we see that the MW model justifies the validity of this postulate and shows that it stems from the quantization condition Eq.\,(16).

However, it is obvious that $T_{cp}=0$ does not mean that the sample temperature has no effect on the properties of the paired electrons, since otherwise $T_c$ would be infinite. Indeed, changing $T$ changes $n_{cp}$\footnote{Loosely,  this is due to the temperature dependence of the polarization of the ionic lattice responsible for the electron pairing.}, as it was shown in the two-fluid model (Eq.\,(1a)). Therefore, the change of $T$ leads to the change of $\lambda_L$ and, correspondingly, to  the change of $r_i$, the rms radius of the field induced motion of electrons in the pairs. On the other hand, $r_i$ is proportional to $R_0$ with the proportionality coefficient determined by the superconducting material, which does not depend on $T$ due to the constancy of $T_{cp}(=0)$. 

Thus, both $r_i$ and $R_0$ depend on the sample temperature in the same way (similar as $\xi_0$ and $\lambda_L$ in the non-local theory of Pippard \cite{Pippard53}). Therefore, for a given material the ratio $r_i/R_0$ is the same as that at $T=0$, and, since both $R_0$ and $r_i$ do not depend on the field, $r_i/R_0$  dependents on neither  $T$ nor $H_0$.  

Below we will use a slightly different ratio: the parameter $\aleph$ (aleph) defined as 
\begin{equation}
	\aleph=\frac{r_i}{R_\perp},
\end{equation}
where $R_\perp$ is the root mean square projection of $R_0$ on the transverse plane  or this is the rms distance (averaged over all possible angles $\theta$) of the orbiting paired electrons from the axis passing through the pair center of mass and parallel to $\textbf{H}$. Important that $r_i$ and $R_\perp$ are radii of concentric circles laying in the same plane.

In Langevin's theory $r_i^2=R_\perp^2=2/3R_a^2$, where $R_a$ is the rms radius of the electron orbits in atom \cite{Tamm}. 
Correspondingly, $\aleph$ is a universal constant equal to 1. Below we will see that in the MW model $\aleph$ is a material constant close in its essence to the GL parameter $\kappa$. 

One more thing. According to thermodynamics, $\Delta S=S_s-S_n(=(V/8\pi)(dH_c^2/dT))$, a difference of entropies of the sample in the MS ($S_s$) and in the N state ($S_n$), does not depend on the field \cite{VK,Shoenberg}. The MW model explains this fact by the complete ordering of the induced magnetic moments in Cooper pairs, similar as it takes place in regular diamagnetics. On the other hand, since the N phase is indifferent to the field by definition, the field independence of $\Delta S$ implies that $n_s$ and therefore $\lambda_L$ does not depend on the field as well (see Appendix for more details). Thus, the field independence of the London penetration depth, following from  the Bohr-Sommerfeld quantization condition in the MW model, agrees with the requirement of thermodynamics, as it should. 

Completing this section,  we note that in the MW model  the penetration depth $\lambda$ is the distance in direction perpendicular to the axis of the  whirls, i.e. to the field $\textbf{H}$. Therefore, in non-cylindrical ellipsoidal samples in the MS the penetration depth in the direction perpendicular to the surface (i.e. to the external field near the surface $\textbf{H}_{ext}$) is equal to  $\lambda \sin\vartheta$, where $\vartheta$ is the angle between $\textbf{H}$ and the normal to the surface $\textbf{n}$ (see Fig.\,2). Hence, this model naturally resolves the aforementioned dilemma of the London theory.    

\section*{FLUX QUANTIZATION} 

Take a sample in the MS, e.g., a cylinder in the  parallel field $H_0\leq H_{c1}$, and consider the quantization condition  Eq.\,(16) but for an arbitrary \textit{macroscopic} closed loop $l$ laying, for simplicity, in a transverse (perpendicular to $\textbf{H}$) plane. Moving along such a loop, we will pass through lots of pairs, so to calculate the circulation over the loop $l$ we should consider the average generalized linear momentum $\langle\widetilde{p}_{cp}\rangle$  and the average quantum number $\langle n\rangle$. Since all Cooper pairs are in identical conditions ($H$ is uniform throughout the sample) $\langle n\rangle =n$. Therefore, 

\begin{equation}
\oint_l \langle\widetilde{\textbf{p}}_{cp}\rangle\cdot d\textbf{l}=nh,
\end{equation}
where $l$ is the loop length.

Now, open $\langle\widetilde{\textbf{p}}_{cp}\rangle$ and take into account that in the MS $n=0$. Then, Eq.\,(46) becomes
\begin{multline}
\oint_l \langle\widetilde{\textbf{p}}_{cp}\rangle\cdot d\textbf{l}=\\\oint_l \langle m\textbf{v}_{i1}\rangle\cdot d\textbf{l}+\oint_l \langle m\textbf{v}_{i2}\rangle\cdot d\textbf{l}+\frac{2e}{c}\oint_l \langle \textbf{A}\rangle \cdot d\textbf{l}=0.
\end{multline}

The first two integrals are zero due to mutual compensation of the induced kinetic liner momenta of electrons in neighboring pairs, like it takes place in the regular diamagnetics. 

Now, what is  $\langle \textbf{A}\rangle$, the average of the vector potential of the $\textbf{H}$-field? To answer, we apply Stokes' theorem; then Eq.\,(47) is rewritten as
\begin{equation}
\frac{2e}{c}\oint_l \langle \textbf{A}\rangle \cdot d\textbf{l}=\frac{2e}{c}\int_{F_l}(\nabla \times\langle \textbf{A}\rangle)\cdot d\textbf{f}=0,
\end{equation}
here $F_l$ is the area of  a surface bounded by the loop $l$ and $d\textbf{f}$ is a vector element of this surface. 

From Eq.\,(48) we see that the integral over the area $F_l$ is the flux of a vector $[\nabla\times\langle \textbf{A}\rangle]$  and this flux equals zero. Therefore, since $\nabla\times \textbf{A}\equiv\textbf{H}\neq0$, $\langle \textbf{\textbf{A}}\rangle\neq \textbf{A}$.

On the other hand, inside our sample the induction $\textbf{B}$ and therefore its flux is zero (Eq.\,(39)). Therefore,  Eq.\,(48) suggests that $\langle \textbf{A}\rangle$ is the vector potential of the magnetic flux density (induction) $\textbf{A}_B$ defined as $\textbf{B}=\nabla\times \textbf{A}_B$. In other words, the vector potential of the flux density 
$\textbf{A}_B$ is a macroscopic average of the vector potential $\textbf{A}$ determining the field induced microscopic currents $J_i$\footnote{Note the exact match with the classical definition of $\textbf{A}_B$  as a macroscopic mean of the vector potential \cite{Tamm}.}.   So, putting $\langle \textbf{A}\rangle=\textbf{A}_B$, we rewrite  Eq.\,(48) as  
\begin{equation}
\frac{2e}{c}\int_{F_l}(\nabla \times\langle \textbf{A}\rangle)\cdot d\textbf{f}=\frac{2e}{c}\int_{F_l} \textbf{B}\cdot d\textbf{f}=\frac{2e}{c}\varPhi=0,
\end{equation}
where $\varPhi$ is the magnetic flux through the area $F_l$.

Now, take a tube-like hollow thick-wall long cylinder, apply the field $H_0(<H_{c1})$ parallel to its longitudinal axis and cool the cylinder below $T_c$. Next, consider the closed loop \textit{l} encircling the cylinder's opening and laying inside the wall far (compare to $\lambda$) from the inner and outer surfaces of the cylinder. The induction inside the wall is zero, implying that, as in Eq.\,(47), $\langle m\textbf{v}_i\rangle=0$. On the other hand, the flux $\varPhi$ inside our hollow cylinder is frozen \cite{Shoenberg} and therefore it is \textit{not zero}. Then Eq.\,(46) yields

\begin{multline}
\oint_l \langle\widetilde{\textbf{p}}_{cp}\rangle\cdot d\textbf{l}=\frac{2e}{c}\oint_l \langle \textbf{A}\rangle \cdot d\textbf{l}=\frac{2e}{c}\int_{F_l}\nabla \times \langle \textbf{A}\rangle\cdot d\textbf{f}=\\\frac{2e}{c}\int_{F_l} \textbf{B}\cdot d\textbf{f}=\frac{2e}{c}\varPhi=nh.
\end{multline} 

Hence, the magnetic flux passing through the opening (plus surrounding it penetration area) in a multiply connected superconductor is 
\begin{equation}
\varPhi=\frac{c}{2e}nh=\frac{\pi c \hbar}{e}n.
\end{equation}

This is the famous London's flux quantization but for the paired electrons. Hence, the origin of the superconducting flux quantization is the quantization condition (16), which also justifies existence of the field induced persistent currents in  the MS, as it \textit{must}. 

Eq.\,(51) indicates that the superconducting flux quantum and therefore the flux passing through \tg{each Abrikosov} vortex in type-II superconductors in the mixed state \cite{VK} is
\begin{equation}
\varPhi_0=\frac{c}{2e}h=\frac{\pi c \hbar}{e}.
\end{equation}  

As well known, the flux quantization (51) and the single flux quantum in the Abrikosov vortices (52) are in full agreement with experiment \cite{Deaver,Essmann,MS}.

Thus, inside the S phase the vector  potential $\textbf{A}_B=\langle\textbf{A}\rangle$, where $\textbf{A}$ is the vector potential of the field $\textbf{H}$.  Now, what is the value of $\textbf{A}_B$ inside the sample in the MS? 

In the plane perpendicular to $\textbf{H}$ we have uniformly distributed identical induced  circular currents, i.e. in-plane currents with the same magnitude $J_{cp}$ and the same rms radius $r_i$ circulating clockwise relative to $\textbf{H}$. Therefore, exactly as it takes place in regular diamagnetics \cite{Tamm}, equal amount of electricity flows in opposite directions throughout \tg{the} out-of-plane cross section of an arbitrary chosen volume element $dV$. Therefore \tg{the} average current density $\langle\textbf{j}\rangle=en_s\langle\textbf{v}_i\rangle$ is zero in any volume element of the sample interior or inside the S phase. Hence, taking into account Eq.\,(22) we obtain 
\begin{equation}
\textbf{A}_B\equiv\langle \textbf{A}\rangle=\langle \textbf{v}_i\rangle=\langle \textbf{j}\rangle=0.
\end{equation}
    
Thus, in the MW model all but one (the incorrect assumption $H_\infty=0$)  assumptions of the London theory (the assumptions in Eq.\,(6) plus (6a) and $\textbf{A}_\infty=0$) follow from the quantization condition Eq.\,(16) provided that $\textbf{j}_\infty$ and $\textbf{A}_\infty$ are average of the corresponding microscopic quantities. At the same time, the highly questionable assumptions in Eq.\,(5) as well as the assumption about circumferential surface current in the MS can be safely ruled out.

\section*{OTHER PROPERTIES} %

\textbf{The Hall effect.} In one of the first experiments on superconductivity, Onnes and Hof revealed a non-existence of the Hall effect in Sn and Pb samples \cite{Onnes-1914}. More specifically, they observed a sharp drop of the Hall voltage  at crossing the critical temperature from above. Later Onnes conducted a carefully designed and long lasting experiment (see footnote ($^1$)), from which he concluded that in superconductors the Hall effect is absent  \cite{Onnes-1924}. These experiments were discussed by Lorentz \cite{Lorentz} and Hall \cite{Hall}; their conclusions are diametrically different\footnote{Lorentz: ``...motion of the electrons is to a great extent insensible to the transverse forces exerted by the field".  Hall: ``... electric currents have in one respect less freedom of motion in the supraconductive state than  in the normal state."}. As of today, no convincing explanation of the non-existence of the Hall effect in superconductors is known.

In the MW model the absence of the Hall effect naturally follows from the fact that the only action of the external magnetic field on the superconducting (paired) electrons (exerted through the field $H$) is the change of magnitude of the angular velocity $\omega$ of their induced circular motion caused by precession of the pairs, regardless on specific path of the transport current. 
Therefore,  the magnetic field cannot affect this path. 
Interesting to note that discussing Onnes' experiments Hall  raised what he called a ``heretical" question: ``Is there conclusive evidence that the persistent currents which Onnes and others have observed are anything more than the aggregate of microscopic electric whirls within the metal?" \cite{Hall}. Needless to say that this is exactly the case in the MW model.

\textbf{Paramagnetism of the Abrikosov vortices.} As  known,  the magnetic moment of type-II superconductors in the mixed state\footnote{Recall that the mixed state  is an equilibrium  (thermodynamic) state \cite{VK}. This implies that magnetization curve of the samples in the mixed state is reversible and correspondingly the samples are pinning free.} being negative increases (decreasing in magnitude) with increasing applied field (see, e.g., \cite{MS,Shubnikov-37}). This implies that each flux line, referred to as Abrikosov's vortex, carries a positive (paramagnetic) magnetic moment. 

The standard picture of the Abrikosov vortices is  shown in Fig.\,8. According to this picture, the induced current in each vortex runs about the vortex core counterclockwise, when viewed from the tip of the field, i.e. it is paramagnetic.  However, as known from electrodynamics \cite{Tamm,Purcell} and thermodynamics \cite{VK} currents induced by magnetic field in a singly connected sample (regardless superconducting or not) are \textit{always} diamagnetic. So the standard picture of the Abrikosov vortices is questionable. On the other hand, numeral experiment confirm the vortex structure of the mixed state \cite{Tinkham,Essmann}. 
\begin{figure}
	\includegraphics[width=0.7\linewidth]{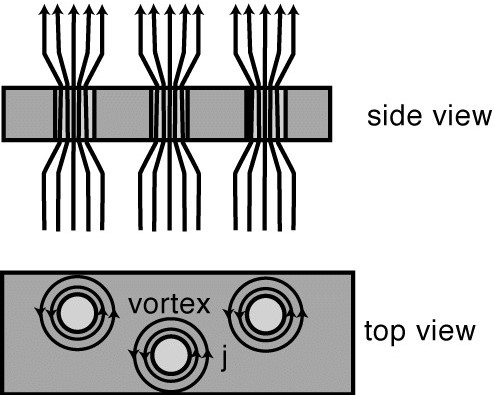}
	\caption{Standard picture of Abrikosov vortices  copied from Google's images of vortices in superconductors.  Persistent current with the volume density $\textbf{j}$ flows counterclockwise inside the field penetration layer along circles centered at the vortex core, i.e. it is paramagnetic. The current charge carriers are Cooper pairs not experiencing resistance. As one can notice the Lorentz force driving the pairs is directed out from the center, and therefore such a current is possible only if either  the charge of Cooper pairs is positive or their mass is negative. Alternatively, this current increases the field passing through the vortex, and therefore it increases the sample free energy, which is contrary to thermodynamics.    } 
	\label{fig:epsart}
\end{figure}

Schematics of the Abrikosov vortex in the MW model is shown in Fig.\,9. In this model the minimum magnetic flux $\varPhi_0$ passes through the sample when superconductivity is suppressed in the volume of one micro-whirl. Thus, there will be a normally conducting ``hole" (not-colored circle) in the network of the superconducting micro-whirls (shown in blue).  
 
Current around this  normal hole (the core of Abrikosov's vortex) is an effective paramagnetic current (shown by the dashed line) formed by the induced diamagnetic currents in precessing Cooper pairs. Correspondingly, contribution of each such a hole into the sample magnetic moment is paramagnetic in agreement with experiment and general physics. 

One can also note that due to symmetry, in sufficiently pure samples the first Abrikosov vortex (the vortex at $H=H_{c1}$) should appear close to the sample geometrical center in the plane transverse to $\textbf{H}$. This has indeed been observed \cite{Zeldov}. 

 On the other hand, in the case of a multi-flux-quantum N domain in type-I superconductors or a set of the single-flux-quantum vortices in type-II ones, the total  flux passing through the sample is an integer of the flux associated with one hole, i.e., $\varPhi_0$. This is an alternative justification of the flux quantization in superconductors.  
\begin{figure}
	\includegraphics[width=0.6\linewidth]{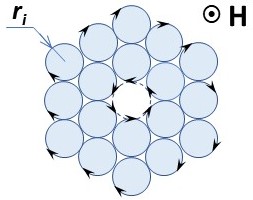}
	\caption{Abrikosov's vortex in the MW model. Blue solid circles with arrows represent the current in the cross section of the micro-whirls (solenoids) colored in light-blue;   $r_i$ is the rms radius of the induced currents. A central not colored circle, the core of the Abrikosov vortex, is a hole in the network of the diamagnetic micro-whirls. An effective paramagnetic current surroundings the core is  shown by the dashed line. It is formed by the diamagnetic currents induced in the pairs surrounding the core.} 
	\label{fig:epsart}
\end{figure}

\textbf{Surface tension}. The most important equilibrium properties of superconductors in the states other than the MS are associated with the energy of the S/N interphase boundaries inside the sample, also referred to as the S/N surface tension. 

As known \cite{Pippard_51,Tinkham,VK}, to construct the S/N surface tension a theoretical model should have two microscopic parameters with dimension of length. In the MW model such parameters are $R_0$ and $r_i$. The wall-energy parameter $\delta$ \cite{Tinkham,VK} in this model is $\delta= (R_\perp - r_i)$. 

In Eq.\,(45) we introduced the parameter $\aleph$. The only what we know so far about $\aleph$ is that it is positive. On the other hand, it is clear that properties of superconductors with $\aleph$ greater and lesser than unity are different. Let us briefly look at what this difference is.

For simplicity, we will again consider the cylindrical sample. Cross sections of the current structure of the MS in the MW model for samples with $\aleph<1$ and $\aleph>1$ are schematically shown in Fig.\,10a and 10b, respectively.  The pink areas there (those with radius $R_\perp$) represent cross sections of the cylindrical volumes filled with Cooper pairs. And the blue areas are cross-sections of the cylindrical volumes (with radius $r_i$) of the micro-whirls/solenoids. We remind that both $r_i$ and $R_\perp$ are root mean square radii, i.e. the edges of the cylinders in Fig.\,10 are not sharp implying that the cylinders overlap not leaving voids.

The field can pass through the sample via the hole in the network of the induced micro-whirls. An elementary hole (the one carrying the single flux quantum $\varPhi_0$\footnote{The flux associated with each elementary hole is the flux passing through an ``empty" blue cylinder  plus an area around it where $B$ decays from $B=H$ to zero.}) represents an ``opening" in the transverse cross section of the 3D network of these currents. The hole volume equals the volume of the \textit{blue} cylinder $\pi r_i^2L$, where $L$ is the length of our sample.  To create such a hole superconductivity must be suppressed in the corresponding \textit{pink} cylinder (the one with radius $R_\perp$). By definition, the magnitude of the minimal field $H$ when it happens is $H_{c1}$.

Now, referring to Fig.\,10a, the magnetic energy of the elementary hole $\Theta_n$, where the subscript $n$ stands for ``normal",  at $H=H_{c1}$ is 
\begin{multline}
\Theta_n=\pi R_\perp^2L\frac{H_{c1}^2}{8\pi}=\\\pi r_i^2L\frac{H_{c1}^2}{8\pi}+\pi(R_\perp-r_i)(R_\perp+r_i)L\frac{H_{c1}^2}{8\pi}.
\end{multline}

On the other hand, in the absence of a hole its space is taken by the cylinder, which magnetic energy  $\Theta_s$ (the subscript $s$ stands for ``superconducting') at $H=H_{c1}$ is
\begin{equation}
\Theta_s=\pi r_i^2L\frac{H_{c1}^2}{8\pi}.
\end{equation}

Therefore, the difference $\varGamma$ of the magnetic energies of the non-superconducting hole and the superconducting ``insert" into this hole is
\begin{multline}
\varGamma\equiv\Theta_n-\Theta_s=\pi(R_\perp-r_i)(R_\perp+r_i)L\frac{H_{c1}^2}{8\pi}=\\\delta bL\frac{H_{c1}^2}{8\pi}=\delta \frac{H_{c1}^2}{8\pi}A_b=\gamma_bA_b,
\end{multline}
where $\delta=R_\perp-r_i$, the radial  width of the space between the pink and blue cylinders, is the wall energy parameter \cite{Tinkham,VK}; $b=2\pi[(R_\perp+r_i)/2]$ is the length of a notional S/N interphase boundary in the plane perpendicular to $\textbf{H}$; and $A_b= bL$ is the area of this boundary. 

We see that $\varGamma$ is the excess energy caused by the presence of the S/N interface at the hole boundary and therefore $\gamma_b=\varGamma/A_b$, the excess energy per unit area of this boundary, is the surface tension. Note that this is close to that how the S/N surface tension was for the first time introduced by H. London \cite{H_London-35} and used by Landau in his laminar 
model of the intermediate state \cite{Landau_37}. On the other hand, the wall energy parameter $\delta=R_\perp-r_i$ is close to that proposed by Pippard: $\delta=\xi-\lambda_L$  \cite{Pippard_51}. 

Now, the total free energy (see \cite{VK}) of the sample with one ``hole" (the N domain carrying the  single flux-quantum) is 
\begin{multline}
\widetilde{F}(T, H_0)_{1h}\equiv\widetilde{F}_{s0}-\int_{0}^{H_{c1}}MdH_0=\\\widetilde{F}_n-\frac{H_c^2}{8\pi}V+(N_\perp-1)\Theta_s+\Theta_n=\\ [\widetilde{F}_n-\frac{H_c^2}{8\pi}V+N_\perp\Theta_s]+\varGamma=\widetilde{F}(T, H_{c1})+\varGamma, 
\end{multline}
where $\widetilde{F}(T, H_{c1})$ is the total free energy of the sample without the hole at $H_0=H_{c1}$, i.e. of the sample in the MS, and $N_\perp$ is the number of micro-whirls (see Eq.\,(42)). 

Thus, we see that if $\aleph<1$, or $\delta>0$, or the surface tension $\gamma_b$ and, correspondingly, $\Gamma$ is positive, the free energy of the cylindrical sample without the hole (N domain) is always (i.e. for any $H_{c1}$) less than that with the hole. Therefore, such a sample stays in the MS all the way up to the thermodynamic critical field $H_c$, or $H_{c1}=H_c$. At $H_0=H_c$ the magnetic energy of this sample $E_m=H_c^2V/8\pi= E_c$ and therefore  superconductivity collapses all over the sample volume meaning that the sample converts to the N state via the phase transition of the first order. Therefore, in the MW model materials with $\aleph<1$  represent type-I superconductors, as it should be the case when the S/N surface tension is positive \cite{VK}. 

\begin{figure}
	\includegraphics[width=0.8\linewidth]{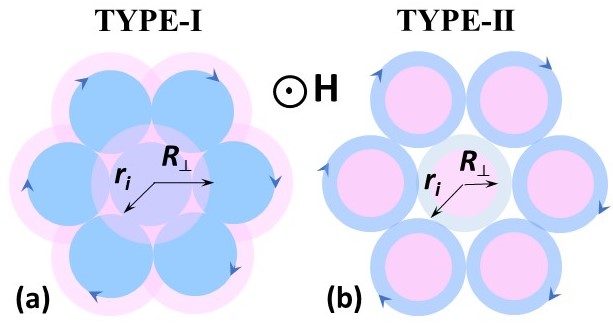}
	\caption{Schematics of the current structure in the transverse cross section of the samples of type-I (a) and type-II (b) superconductors. In (a) $r_i<R_\perp$ or $\aleph<1$; in (b) $r_i>R_\perp$ or $\aleph>1$. The induced currents are designated by arrows.  Areas filled with Cooper pairs are colored in pink; the cross-sectional areas of the induced micro-whirls are colored in blue. $R_\perp$ is the projection of the rms radius of the orbital motion of the paired electrons $R_0$ onto the plane perpendicular to $\textbf{H}$; $r_i$ is the rms radius of the induced motion of the paired electrons. The field $\textbf{H}$ is directed toward the reader.} 
	\label{fig:epsart}
\end{figure}

The same steps as above for the case depicted in Fig.\,10b where $R_\perp<r_i$ lead to Eq.\,(56), but  now $\varGamma<0$, or the surface tension is negative, or $\aleph>1$. Then from Eq.\,(57) we see that the free energy of the sample with the hole is \textit{always} less than that without the hole. This means that the sample should be in the mixed state at any $H_0$ regardless how small this field is. Apparently, however, that the flux quantization prevents the appearing of the first hole until $H(=H_0$ in the cylindrical sample) reaches a finite value $H_{c1}$. At this field  the flux passing through the first hole equals $\varPhi_0$. The same conclusion can be drawn from purely thermodynamic reasoning \cite{VK}. 

To create the second hole, the applied field $H_0(=H$ in our sample) should be increased so that the total flux passing through the sample equals $2\varPhi_0$, and so on. At the same time, it is easy to show that separate holes with the flux $\varPhi_0$ passing through each are thermodynamically more profitable than one hole with the total flux in it \cite{LP,Tinkham}. As known, this picture is fully consistent with the experiment (see, e.g., \cite{MS}). 

Therefore, above $H_{c1}$ the number of holes (Abrikosov vortices or the flux lines) gradually increases with increasing $H_0$ and the transition to the N state is continuous (second order) phase transition occurring at  $H_0=H_{c2}>H_c$. In the MW model $H_{c2}$ is\footnote{Eq.\,(54) is derived assuming that superconductivity vanishes completely at $H_{c2}$, i.e. neglecting the filamentary state occurring at $H_{c2}<H_0<H_{c3}$ \cite{VK,Filaments}.}
\begin{equation}
H_{c2}=\aleph H_c.
\end{equation}

Thus, we see that materials with $\aleph>1$ represent type-II superconductors and the significance of the parameter $\aleph$ is similar to that of the GL parameter $\kappa$. We remind that  the range of applicability of the GL theory is limited to the close vicinity of the critical temperature \cite{Gorkov}.

Applying thermodynamics and taking into account that the governing field inside superconductors is the field intensity $\textbf{H}$ one can extend the above consideration to arbitrary  ellipsoidal samples, i.e., to both homogeneous (Meissner) and non-homogeneous (intermediate and mixed) states. 

\textbf{Type-I to type-II conversion. } As known, the addition of an impurity to a flawless elementary type-I superconductor (e.g., alloying with another metal) converts it to a type-II superconductor and, correspondingly, leads to an increase of the critical field of the S/N transition denoting in this case as  $H_{c2}$  \tg{\cite{Shubnikov-37}}. The same effect takes place at decreasing  the sample dimensions, in particular the thickness of sufficiently  thin films (see, e.g. \cite{Shoenberg,Tinkham} and references therein). Shubnikov et al.\footnote{In this work the type-II superconductivity was solidly established for the first time.} \cite{Shubnikov-37} reported that the higher the alloying percentage, the higher $H_ {c2}$.  And Pippard reported \tg{\cite{Pippard53}} that the alloying increases $\lambda_L$ and reduces the mean free path length.  

Pippard  explained the type-I/type-II conversion by decreasing the coherence length $\xi(=(1/\xi_0+1/l_m)^{-1})$, assumed to be a function of the mean free path $l_m$. Recall that according to Pippard $\delta=\xi-\lambda_L$, and $\xi$ is a spacial scale over which a change of F. London's ordering of the electron structure with the rigid $\widetilde{\textbf{p}}$ can take place (``spreading tendency of the state of order" \cite{Pippard_87}). Correspondingly, the decreasing mean free path (which can also be expected in thin films) results in decreasing $\xi$ and therefore can lead to the sign change of $\delta$ \cite{Pippard53}.   

For the thin films an alternative explanations of the increase of the critical field  was given by H. London \cite{H_London-35} and by Ginzburg and Landau near $T_c$ \cite{GL}. 

In the MW model $\xi(=2R_0)$ is the size of Cooper pair, a temperature-dependent constant of material determined \tg{by a polarization}  ability of its ionic lattice. At alloying, atoms of the other metal locally disturb the polarization of the lattice of the host material thus preventing the formation of pairs, which otherwise would exist at that locations. Apart from these places, the polarization stays unchanged. Correspondingly, the number density  of Cooper pairs $n_{cp}$ decreases, but $\xi$ is not altered. This leads to increase of $\lambda_L$, $r_i$ and $\aleph$. The latter, being originally  lesser than unity, at sufficient alloying (in PbTl alloy it can be as small as 0.8\% of Tl \cite{Shubnikov-37}) becomes greater then one, meaning that the sample material converts from type-I to type-II.

This picture is consistent with Pippard's observation of the increase of  penetration depth at the alloying and with the fact that pure type-I superconductors can be converted to type-II ones, but  not vice versa\footnote{Type-I to type-II conversion can be also achieved by introducing structural defects. In such case the opposite conversion is possible by annealing.}.

A similar in its essence process leads to an effective type-I/type-II conversion of the film material at decreasing the film thickness $d$. When $d$ becomes less than about $2R_0$ only Cooper pairs oriented so that $2R_0$\,sin$\phi<d$, where $\phi$ is the angle between $\vec{\mu}_0$ and the normal to the film surface, can survive. Respectively, the lesser $d$ the lesser $n_{cp}$ and the greater  $\aleph$. This implies that at a definite thickness $\aleph$ becomes larger than unity and therefore the film behaves as though it is made of \tr{a} type II superconductor.  Correspondingly, with a further decrease of the thickness the critical field becomes progressively greater than $H_c$.

The outlined interpretation is consistent with the well known fact that even very thin films of type-I materials remain superconducting and that these films always behave as type-II superconductors, as for the first time was revealed in the classical experiment of Shalnikov \cite{Shalnikov} and confirmed in many other experiments afterwards.   

\textbf{Total current.} Finally, let us briefly consider one more very important property of superconductors not directly related to the MS, a so called total current. 

In the MW model the equilibrium magnetic properties of superconductors are qualitatively similar to the properties of conventional diamagnetics. In both cases these properties are due to magnetization arising from precession of the microscopic magnetic moments caused by the orbital motion of electrons bound either in atoms (conventional diamagnetics) or in Cooper pairs (superconductors). A colossal quantitative  difference in the magnetic susceptibilities in these materials (up to 5 orders of magnitude!) is due to the differences in the size of the orbits and correspondingly in the radii of the induced microscopic currents of the bound electrons. 

However, there is also an important qualitative difference in these induced currents. Namely, in conventional diamagnetics the orbiting electrons are bound in motionless atoms (fixed in the crystal lattice), while in superconductors - in movable Cooper pairs. Therefore, since the pairs have electric charge, they can form an electric current, referred to as the total current \cite{Shoenberg}. It includes the transport current and the field-induced current encircling the opening in multiply connected bodies\footnote{The current in closed circuits, like, e.g., superconducting magnets, as well as the current encircling the flux trapped in so called pinning centers in insufficiently pure superconductors has the same nature as the total current in a superconducting ring. A perfect discussion of the latter is available in \cite{Shoenberg}.}. In the latter case, as is observed in the experiment \cite{Deaver}, the total current is quantized due to the flux quantization.

The total current in superconductors plays a role similar to that of the transport current  in conventional metals, where it is executed by conduction electrons. The latter obey Fermi-Dirac statistics and their energy equals the Fermi energy $E_F$. This implies that the carriers of the transport current in normal metals, or in the N phase of superconductors, are  "very hot": their temperature (the temperature of the ensemble of conduction electrons) equals the Fermi temperature $T_F= E_F/k_B\sim 10^4$ K, where $k_B$ is the Boltzmann constant \cite{Kittel}. 

In striking contrast, the charge carriers (Cooper pairs) of the total current in the S phase are "deadly cold". Since their  spins are zero, the pairs obey Bose-Einstein statistics and, since the temperature of the ensemble of  the paired electrons $T_{cp}$ is zero, they form the Bose-Einstein condensate (BEC).  This (zero temperature) naturally leads to the disappearance of the thermoelectric effects, as observed in experiments (see \cite{Shoenberg} for references). This amazing transformation of the ``very hot" single electrons to the ``zero-temperature" electron pairs can be compared with the fact known from the relativity theory: two flying apart massless photons form a  massive pair located in their center of mass \cite{Okun}.

Therefore, the total current in superconductors represents the transport current in BEC, known on the properties of superfluid helium (see, e.g., \cite{LP}). Correspondingly, the total current set in a closed superconducting circuit continues running without energy dissipation provided speed of the charge carriers (density of the total current) is lesser of a definite critical value.

Thus, in the MW model the electromagnetic properties of superconductors associated with the total current\footnote{We are talking about dc total current. Consideration of the ac current must include contribution of the unpaired electrons, which is significant at frequencies $\gtrsim10^3$ MHz \cite{Shoenberg,Tinkham}. }  resemble  the properties of hypothetical perfect conductors. 
 
As mentioned, the hell-mark of the perfect conductors is irreversibly of the sample magnetic moment induced by the changing applied magnetic field  \cite{VK,Shoenberg}; its direction is determined by the Lenz law whereas the magnetic moment caused by magnetization is always diamagnetic. One more important feature distinguishing the bound and total currents in superconductors is that energy of the former comes from internal resources, i.e., condensation energy \cite{VK}, whereas energy of the latter is supplied by an external source\footnote{This can be a battery, which is needed to maintain the transport current in the non-superconducting parts of the circuit and/or to set the current in the superconducting magnet, or e.m.f. caused by a change of the applied magnetic field in the case of multiply connected bodies.}. For this reason the multiply connected superconductors can never be in the thermodynamic equilibrium implying that for such bodies the principle of the free energy minimum is inapplicable.

However, it should be stressed that even in the presence of the total current, magnetic properties of superconductors in the MW model are different from those of the perfect conductors. Since the total current is accompanied by its own magnetic field, it represents a combination of the whirl and  translational motion of the paired electrons regardless of the presence or absence of the applied field. 
So in the MW model the total current is, in fact, similar to the superfluid current in He-II \cite{Feynman57}. This is consistent with magnetic properties of the superconducting ring\footnote{Interpreting the magnetization curve of the ring Shoenberg distinguishes the total current and what he calls "surface" current.} \cite{Shoenberg} and with the aforementioned observation of Meissner and Ochsenfeld \tr{in} the fourth arrangement of the samples, namely ``When the parallel superconductors are connected end-to-end in series and an external current is connected to flow through them above the critical temperature the magnetic field between the superconductors is increased below the transition temperature the external current being unchanged." \cite{Meissner}.

\section{EXPERIMENT}

Above we have already mentioned sufficiently many experimental facts  supporting the  MW model. Some of them, being known for many decades, are explained in the new model for the first time. In this section we will briefly stop at  yet preliminary results of our recent Low Energy muon Spin Rotation (LE-$\mu$SR) experiment
 \cite{muons-18}. The endeavor to understand these results was the immediate reason of the appearance of the presented model.   

This experiment was targeted to verify a possible field dependence of the penetration depth near the sample surface. Such a possibility follows (i) from the GL theory, where the field dependence of the order parameter is the main  feature distinguishing  the electromagnetic properties of the MS from those in the London theory; (ii) from reported $\mu$SR and SANS (Small-Angle Neutron Scattering) data obtained on the samples in the mixed state and treated using the GL and London theories; (iii) from the standard diagram of the Abrikosov vortices; and (iv) from a thermodynamic argument of H. London following from the assumption of the circumferential screening current of the London theory (see Appendix). Earlier $\lambda(H_0)$ was measured by Pippard using a microwave resonator; Pippard concluded that $\lambda$ can be considered as independent of $H_0$ \cite{Pippard50}. Later Sridhar and Mercereau demonstrated \cite{Sridhar} that Pippard's data could be subject to non-equilibrium effects caused by the high-frequency radiation, although the magnitude of these effects is not clear (see footnote ($^{48}$)). 

So we set up an experiment to  verify  the field dependence of the penetration depth near the surface of samples in the MS with $\eta=0$ using the LE-$\mu$SR spectroscopy. The samples were very pure In and Nb films\footnote{See \cite{MS,IS-3} for characterizations of the films used.} with the thickness about 3 $\mu$m each in the field $H_0$ applied parallel to the films.

\begin{figure}
	\includegraphics[width=0.8\linewidth]{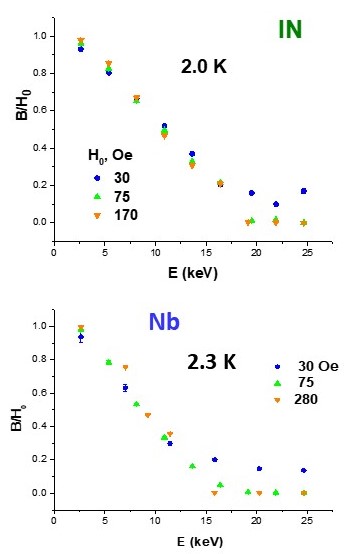}
	\caption{Induction profile near the surface of the indium (upper panel) and niobium films at indicated values of the applied field directed parallel to the film. $E$ is kinetic energy of the implanted muons; an average implantation depth is directly proportional to $E$.  } 
	\label{fig:epsart}
\end{figure}

Data for the induction near the samples surface obtained at temperature near 2 K are shown in Fig.\,11. For the niobium sample similar data were obtained also at 8 K. As one can see, the induction profiles do not depend on the applied field\footnote{In both films there is a strange deviation from the common trend of the B-points measured by high-keV muons at $H_0$ = 30 Oe. Similar deviation takes place in the B-data for Nb at 8 K. This may be an artifact (e.g., due to a frozen-in Earth field) requiring verification.} and therefore the penetration depth $\lambda$ is field independent  in both samples. This result is consistent with Pippard's data \cite{Pippard50}. The fact of the field independence of $\lambda$ is inline with the London theory \cite{London50,London35} and with the MW model. 

Important to stress that in both samples for all  $H_0$ (but 30 Oe) $B$ measured with high-keV muons is zero (it is not so in Nb at 8 K). Specifically, in the In sample $B=0$ when measured using muons with the energy  $E\gtrsim$ 20 keV, which corresponds to the average muon implantation depth $z\gtrsim$ 90 nm. In the Nb sample $B=0$ when $E\gtrsim$ 16 keV ($z\gtrsim$ 50 nm). This indicates that in both samples $B$ is homogeneous in a statistically significant number of the sites of muons with energy $E=24.6$ keV, the maximal energy of muons in this experiment. In Fig.\,12 we show the muon stopping profiles in Nb (see \cite{IS-3} for In). As one can see, at low temperature nearly all muons with $E=24.6$ keV stop at the depths where the induction $B(=0)$ is homogeneous. 
Therefore, like in the normal state, the stopping distribution of these muons is inconsequential for the data retrieved  from their spectra.

\begin{figure}
	\includegraphics[width=0.9\linewidth]{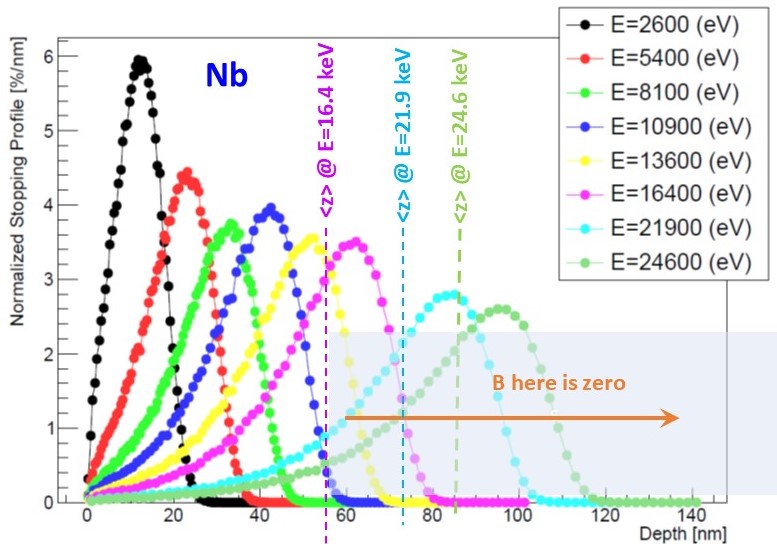}
	\caption{\label{fig:epsart} The stopping distributions of muons of different energies $E$ in niobium, as calculated from the Monte-Carlo code \texttt{TRIM.SP}. The vertical dashed lines denote the mean stopping distances at corresponding energies. A shadowed area designates the depths range where $B=0$ at $T=$ 2.3 K.   }
\end{figure}

\begin{figure}
	\includegraphics[width=0.8\linewidth]{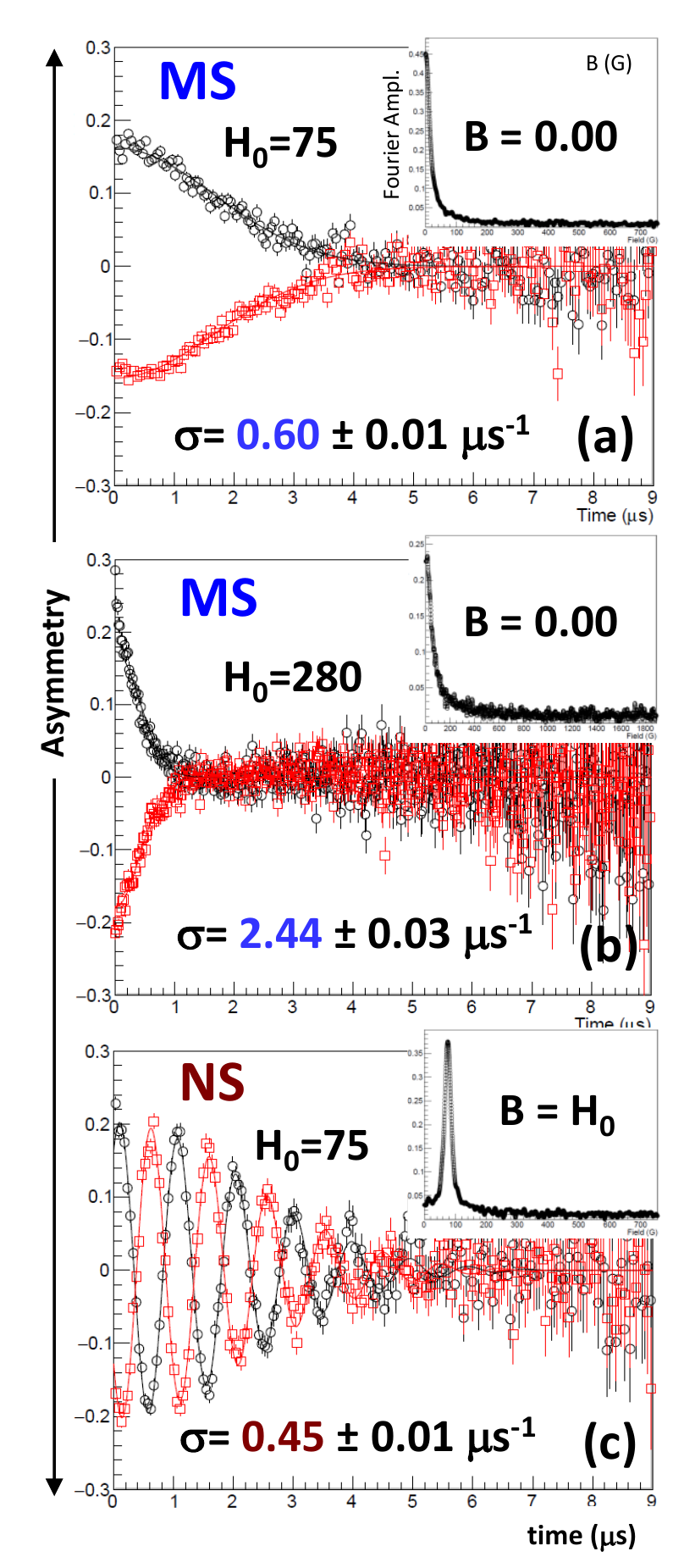}
	\caption{LE-$\mu$SR time spectra taken on the Nb sample. The Fourier transforms of these spectra, representing the microscopic field distributions over the muon sites, are shown in the inserts.  The spectra in (a) and (b) were obtained on the sample in the Meissner state (MS) at $T=2.3$ K using muons with energy $E=24.65$ keV. The spectra in (c) were obtained on the sample in the normal state (NS) at $T=11$ K. $H_0$ is the applied field in Oe; $\sigma$ is the depolarization rate. See \cite{IS-3} for methodical details about the measured spectra. } 
	\label{fig:epsart}
\end{figure}

\begin{figure}
	\includegraphics[width=0.9\linewidth]{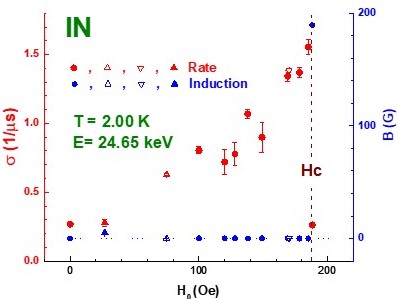}
	\caption{The muons' depolarization rate (red symbols) and the induction (blue symbols) versus the applied field $H_0$ in  the In film at temperature 2.0 K (the film is in the MS) measured with muons of energy 24.65 keV. Corresponding average implantation depth is about 120 nm. Solid circuits are the data obtained in field scans with muons of fixed energy;  triangles are the data obtained in energy scans at fixed fields 30 Oe (solid triangle), 75 Oe (open triangle up) and 170 Oe (open triangle down). $H_c$ is the critical field.} 
	\label{fig:epsart}
\end{figure}

On the other hand, if it is so, the spectra of the microscopic fields probed by these muons (again, like in the \tr{N} state) should be Gaussian. In Fig.\,13 we show the time spectra taken on the Nb sample using the high-keV muons and the spectra taken 
in the \tr{N} state (at $T=11$ K). The  Fourier transforms of these spectra, shown in the inserts, represent the spectra of  microscopic fields. We see that these spectra are Gaussian in \textit{all three} data-points. This confirms that the stopping distribution of the high-keV muons does not affect the data obtained on this sample at low temperature. Similar conclusion can be drawn for the  In sample.

After all, the Gaussian spectra of the microscopic fields implies that the Gaussian approximation, which was used to fit the spectra, is completely adequate and therefore the high-keV muons deliver correct information not only on $B$ (as they always do \cite{Andreas}) but also on the muon depolarization rate $\sigma$, the damping coefficient of the $\mu$SR signal. The latter characterizes the inhomogeneity of the microscopic field over the muon sites (see \cite{IS-3} for details).

Data for the induction $B$ and for the rate $\sigma$ in the pure S phase (i.e. at the depths where $B=0$) for In sample are shown in Fig.\,14. These data provide the most direct support for the MW model.

The point is that, as was just shown, in both our samples at low temperature a statistically significant fraction of muons with $E=24.6$ keV  stop  beyond the penetration layer. According to the standard theories, at these depths the superconducting material is totally inert (no fields, no currents). If so, the muon depolarization rate must be the same as $\sigma(0)$, the rate in the S phase at $H_0=0$. The latter should be close to $\sigma_N$, the rate in the N state, since $\sigma(0)$ and $\sigma_N$ are mostly determined by the nuclear damping. Indeed, as seen from Fig.\,14, $\sigma(0)\approx\sigma_N$ (extreme left and right red data-points). 

On the other hand, according to the standard theories at $0<H_0<H_{c1}$ the rate should be the same as $\sigma(0)$, i.e. it \textit{should not} depend on $H_0$ because the latter is supposed to be completely screened at that depth. The situation is right opposite in the MW model: the greater $H_0$, the greater $H(=H_0$ in the given geometry) and therefore the greater the induced currents $J_i$ circulating all over the sample volume. Correspondingly, inhomogeneity of the microscopic fields increases with increasing $H$. If so, the depolarization rate of muons stopped at the depths where $B=0$ should be increasing with increasing $H_0$. This is exactly what we see in Figs.\,13a and b, and 14.

\section*{SUMMARY AND OUTLOOK}

Historical experiments of Meissner and Ochsenfeld and of Rjabinin and Shubnikov, in which the Meissner effect was established, are reviewed. Theories of Gorter and Casimir and of F. and H. London addressing, respectively, thermodynamic and electromagnetic properties of superconductors in the Meissner state are reviewed as well. It is shown that due to significant difficulties in Londons' theory, the standard description of the electromagnetic properties of superconductors is inadequate and requires revision.  

A novel semi-classical micro-whirls model targeted to address this issue is developed and presented. The model is based on the concept of paired electrons obeying the Bohr-Sommerfeld quantization condition.  Accordingly, each Cooper pair represents a microscopic circular current like the current due to orbiting electrons in atoms. 

The model is valid in the whole range of the superconducting state for samples of any shape and at any orientation of the applied field. A fundamentally new prediction of the model is that the S phase of superconducting ellipsoidal bodies in a static magnetic field is filled with ordered microscopic whirls of field induced currents caused by precessing Cooper pairs. The model is free from disadvantages of the standard theories, simultaneously reproducing their achievements. All properties predicting by the model can be verified experimentally.  

The model consistently describes equilibrium properties of superconductors, which includes the Meissner  effect, persistency of induced currents, zero entropy of ``superconducting" electrons (i.e. of the ensemble of Cooper pairs), the London rigidity principle for Cooper pairs,  the flux quantization,  the  S/N interface energy of two signs and hence two type of superconductivity,  and others. Some well-known experimental facts, such as the Meissner state in non-spheroidal samples and paramagnetism of the Abrikosov vortices, are consecutively described in the new model for the first time. Non-equilibrium properties of superconductors, e.g., the non-existence of the Hall effect and other properties associated with the total current are explained by the model as well.

According to the new model, there are two kinds of non-dissipating currents in superconductors. 
(1) The current formed by electrons bound in stationary Cooper pairs; it arises 
due to precession of the magnetic moments of the pairs caused by the orbital motion of the coupled electrons about their center of mass; these microscopic currents dictate thermodynamic (equilibrium) properties of the singly connected isolated ellipsoidal samples, i.e., properties of the Meissner, intermedium and mixed states.  (2) A dissipation-free flow of Cooper pairs forming the total current leading to the non-equilibrium properties resembling the properties of hypothetical perfect conductors. Both equilibrium and non-equilibrium properties follow from the single Bohr-Sommerfeld quantization condition for the paired electrons (Eq.\,(16)). 

There was a long lasting discussion about what is the prime property of superconductors, i.e. of the superconducting fraction of conduction electrons or the ensemble of Cooper pairs, either  it is zero resistivity or zero induction \cite{Landafshitz_II,DeGennes,Shoenberg}. For some reason, this discussion often overlooks the third fundamental property postulated by Gorter and Casimir: zero entropy \cite{Gorter_Casimir}. In the micro-whirls model these three  key characteristics of superconductivity are on an equal footing and share the same root: quantization of the orbital motion of the  paired electrons. 
 
Contrarily to the standard theories, the model predicts a strong field-dependent inhomogeneity of the induced microscopic currents in the bulk of the samples in the Meissner state. The existence of such currents follows from the requirement of the First law of thermodynamics and were foreseen by Hall \cite{Hall} coming from the disappearance of the Hall effect first observed by Onnes and Hof \cite{Onnes-1914}, and by I. Kikoin \cite{Isaak-2} based on his and Goobar measurements of the gyromagnetic ratio \cite{Isaak-1}. Available LE-$\mu$SR data support this prediction. However these data were obtained in the experiment aimed to another task and therefore require a targeted verification.  It can be also interesting to perform the  SANS experiment with samples in the Meissner state. Potentially, such an experiment can make it possible to measure the diameter of micro-whirls. 

One more way to verify the model can be as follows. According to the model, the sample in the Meissner state is filled with the ordered and parallel to $\textbf{H}$ whirls of the field-induced microscopic circular currents with a frequency $\omega=eH/2mc$, where $\textbf{H}$ is the field intensity inside the sample  (Eq.\,(30)). Therefore, if it is so, a weak ac transverse field (with respect to $\textbf{H}$) should experience a resonance absorption at the linear frequency $\nu_s=\omega/2\pi=1.8H$ MHz, where $H$ is in Oe (numerically the same as G). The frequency $\nu_s$ equals half the standard frequency of the  electron paramagnetic resonance (EPR). Important to stress that the sample \textit{should not be small}. For example, it can be a piece of straight superconducting wire with a diameter of 0.5 mm and a length of 5-10 mm or longer. In such case $\textbf{H}=\textbf{H}_0$ if the applied field $\textbf{H}_0$ is parallel to the sample longitudinal axis, and $\textbf{H}=2\textbf{H}_0$, if $\textbf{H}_0$ is perpendicular to this axis. For a spherical sample (e.g., 4 mm in diameter like the sample used in the experiment of I. Kikoin and Goobar) $\textbf{H}=3\textbf{H}_0/2$. To ensure unambiguity, the sample should be pure and free of frozen-in flux.

Note that a possible existence of such a diamagnetic resonance can be traced from measurements of the gyromagnetic ratio in superconducting samples by I. Kikoin and Goobar \cite{Isaak-1,Isaak-2} similar as the existence of  EPR is traced from measurements of the gyromagnetic ratio in ferromagnetic samples by Einstein and de Haas\footnote{In both works the gyromagnetic ratio was measured using a resonance technique, in which the sample oscillates in resonance with the oscillating applied field.} \cite{Einstein}. 
 
Regarding the model itself, its story is not complete yet, of course. One of remaining questions is the penetration depth at the surface of the samples in the MS and at the S/N interphase boundaries within the samples in the intermediate and the mixed states. 

Perhaps the main feature of the new model is its striking simplicity. As cited above, a model of such kind was anticipated by Fritz London based on his discovery of the rigidity principle. On the other hand, being semi-classical, the model can not and does not address specifics of the lattice polarization responsible for Cooper pairing.

Finishing, we cite one more excerption from the book of F. London \cite{London50}: "Thus we may conclude that it ought to be \textit{sufficient} [for construction of the model] to derive the rigidity of the average momentum for simply connected isolated superconductor". It seems we have strong reason to say that the presented model copes with this task.

\section*{APPENDIX}
\textbf{Thermodynamic argument of  H. London.}
\vspace{2mm}

In 1947 Heinz London suggested a thermodynamic argument according to which the London penetration depth $\lambda_L$ should depend on the applied field $H_0$ \cite{HLondon47,Pippard50,Shoenberg}. This argument has played an important role in the history of superconductivity, but is rarely mentioned in textbooks. Let us recall and discuss it. 

Consider a long superconducting (for definiteness type-I) rod of radius $\Re\gg \lambda_L$ in a parallel field $\textbf{H}_0$, i.e., a cylindrical sample in the MS.   A strict thermodynamic relationship  reads for this sample\footnote{In general case, i.e., for any ellipsoidal sample in an arbitrary oriented field this relationship is $(\nabla_\textbf{H}S)_T=(\partial \textbf{M}/\partial T)_H$. } (see, e.g., \cite{VK})
\begin{equation}
	\left(\frac{\partial S}{\partial H_0}\right)_T=\left(\frac{\partial M}{\partial T}\right)_{H_0},
\end{equation}
where $S$ and $M$ are the entropy and magnetic moment of the sample, respectively, and $T$ is its temperature.

In the London theory the magnetic moment of the chosen sample is (see footnote ($^{20}$) above)

\begin{equation}
	M=-\frac{H_0}{4\pi}(V-V_p)=-\frac{H_0}{4\pi}(V-\lambda_L A_{sm}),
\end{equation}
where $V_p$ and $A_{sm}$ stand for the volume of penetration layer and the sample surface area, respectively.

Note that $V_p$ represents a kind of ``excluded" volume possessing pretty peculiar properties: it does not contribute into the sample magnetic moment, but the current in  this very volume sets this moment. 

Next, neglecting thermal expansion and taking into account that $\lambda_L$ depends on $T$, from Eq.\,(59) it follows  
\begin{equation}
	\left(\frac{\partial S}{\partial H_0}\right)_T=\frac{A_{sm}H_0}{4\pi}\left(\frac{\partial \lambda_L}{\partial T}\right)_{H_0}.
\end{equation}

Therefore,  since $(\partial\lambda_L/\partial T)_{H_0}\neq 0$ \cite{Shoenberg}, the sample entropy $S$ depends on $H_0$ at constant temperature. Important to point out that this entropy is contained entirely within the penetration layer\footnote{This is another strange property of $V_p$, which is excluded from the production of magnetic moment.} (if $V_p=0$ then there is no the entropy change with the field) 
and its density (calculated as $(S-S_0)/V_p$, where $S_0$ is the entropy at zero field) is \textit{ not small} \cite{Pippard50,London50,Shoenberg}.    

According to the two-fluid model, the entropy of a superconductor in the MS is caused by the change of $n_s$\footnote{Actually, as we have seen, this is not quite a lawful extension of the two-fluid model, where $n_s$ changes  only with temperature.}. Hence, the field dependence of $S$ implies that $n_s$ depends on the field at constant temperature.  Therefore, $\lambda_L(\sim 1/\surd n_s)$  depends on the field. This is the essence of  H. London's argument. Recall that the \textit{field-independence} of $\lambda_L$ is one of the base assumptions of the London theory. So, H. London's argument poses one more dilemma in this theory.

This argument was discussed by Pippard \cite{Pippard50} who, as mentioned above, found experimentally that $\lambda_L$ is  essentially \textit{field-independent}. Pippard's discussion is reproduced in \cite{London50} and \cite{Shoenberg}, therefore here we note only two points. (i) Pippard dropped  from consideration 70\% of his data\footnote{Those are the data at 1.5\,K $\lesssim T\lesssim$ 3\,K. At these temperatures the reported variation of the \tg{effective} penetration depth $\delta \lambda_L=\Delta\lambda_L/\lambda_L$ at $H_0$ changing from zero to $H_c$ decreases from about 0.023 at 1.5\,K to 0.002 at 3\,K. Above 3\,K $\delta \lambda_L$ increases up to 0.027 near $T_c$. From Eq.\,(61) (after two more steps) it follows  that $\delta \lambda_L$ increases with increasing temperature starting from zero at 0\,K. On that ground Pippard concentrated on the data at $T>$ 3\,K.} thus admitting, in fact, that  error bars he provides are not fully adequate. Indeed, as it was demonstrated by Sridhar and Mercereau \cite{Sridhar}, Pippard's data can be subject to non-equilibrium effects caused by high-frequency radiation, which have not been accounted. (ii) Basing on  the remaining data and (although not always consequent) interpretation of  Eq.\,(61) Pippard concluded that F. London's idea of a long-range ordering of the superconducting phase is consistent with his observations. Coming from that Pippard arrived at the concept of nonlocality \cite{Pippard53}, which paved the way for the BCS theory.

Now, let us consider what one can say about $\lambda_L$ based solely on  thermodynamics.   

By definitions of the condensation energy $E_c(T)$ and the thermodynamic critical field $H_c(T)$, the difference between the total free energies of the sample in the superconducting ($\widetilde{F}_s$) and the normal ($\widetilde{F}_n$) states in zero field is\footnote{In zero field $\widetilde{F}(T,\textbf{H}_0)=\hat{F}(T,\textbf{H})=F(T,\textbf{B})$, where the last two functions are the Gibbs and Helmholtz free energies, which are appropriate for samples of the cylindrical and transverse geometries, respectively. The total free energy $\widetilde{F}$ is suitable for all geometries \cite{VK}.}
\begin{equation}
	\widetilde{F}_{n0}-\widetilde{F}_{s0}=\widetilde{F}_n-\widetilde{F}_{s0}= E_c(T)= V\frac{H_c^2(T)}{8\pi},
\end{equation}
where subscript $0$ designates zero field; for the normal state $\widetilde{F}_{n0}=\widetilde{F}_n$ since the magnetic permeability in the normal state is unity by definition. 

Next, by definition of the total free energy \cite{VK}, $\widetilde{F}_s$ of a cylindrical sample in the field $\textbf{H}_0$ at constant $T$ is
\begin{equation}
	\widetilde{F}_s(T,H_0)=\widetilde{F}_{s0}-\int_{0}^{H_0}\textbf{M} d\textbf{H}_0=\widetilde{F}_n-V\frac{H_c^2}{8\pi}+V\frac{H_0^2}{8\pi}.
\end{equation}  

Here we also used the Meissner state definition according to which  $\textbf{M}$ of the sample in the MS does not depend on temperature at constant field, as it takes place in all other diamagnetics \cite{VK}. As known (see, e.g., Fig.\,3), this is consistent with  experiment.   

Now, from Eq.\,(63) we get
\begin{equation}
	S_s\equiv-\left(\frac{\partial\widetilde{F}_s}{\partial T}\right)_{H_0}=S_n+\frac{VH_c}{4\pi}\frac{dH_c}{dT}.
\end{equation}

\begin{figure}
	\includegraphics[width=0.9\linewidth]{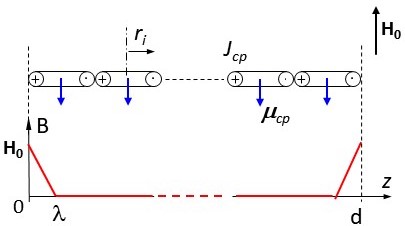}
	\caption{A cross section of the field-induced currents in Cooper pairs $J _{cp}$ inside a thin traverse (with respect to $\textbf{H}$) slice of a sample in the MS as follows from the MW model. The sample is a film with thickness $d$, like the film in Fig.\,3. The slice contains currents induced in one Cooper pair of each micro-whirl of the sample; $\mu_{cp}$ is the magnetic moment induced in  one  pair. The whirls represent identical solenoids parallel to the field $\textbf{H}(=\textbf{H}_0$ in this geometry); their rms radius $r_i=2\lambda_L$.  The graph schematically shows the distribution of the induction $B$ along the $z$ axis perpendicular to the sample surface; $\lambda(\neq \lambda_L)$ is the width of the surface layer in which $B$ decays from $H_0$ to zero. } 
	\label{fig:epsart}
\end{figure}

Since none of the terms on the right hand side depend on the field, the entropy of the superconducting sample in the MS does not depend on the field (like in all other diamagnetics). This implies that $n_s$ and therefore $\lambda_L$ are \textit{independent} of the field. The same conclusion can be drawn immediately from Eq.\,(59) basing on the  MS definition. Following to Pippard \cite{Pippard50}, the field independence of $\lambda_L$ means that the long-range order of the superconducting phase extends over the entire volume of the sample in the MS. Needless to say that this is exactly what follows from the MW model.   

Thus, thermodynamics applied to the London theory requires the field dependence of $\lambda_L$, the agreement of which with experiment is, at least, questionable.  On the other hand, in accordance with experiment, the pure thermodynamic approach excludes such a dependence. However, it may look like that the latter approach ignores the presence of the penetration layer where $B\neq0$, which is confirmed in countless experiments. Let us now turn to the MW model.  

Fig.\,15 schematically shows a cross section of a thin ($\approx$ 6 fm thick) transverse slice of a sample in the MS, as it follows from the MW model. The slice contains the induced currents  in one Cooper pair $J_{cp}$ of each micro-whirl of the sample.

As one can see, contrarily to the London theory, there is no excluded volume. The currents induced in a unit volume just next to the sample boundary (at $z$ near $0$ and $d$) make the same contribution to the sample magnetic moment as the induced currents in the unit volume in any other part of the sample. Therefore, the magnetic moment per unit volume $\chi$ is the same everywhere inside the sample. Then, the sample magnetic moment is 
\begin{equation*}
	M=\chi H V=-\frac{1}{4\pi}HV=-\frac{1}{4\pi}\frac{H_0}{(1-\eta)}V,
\end{equation*}
where $H$ is the field intensity, $\eta$ is the demagnetizing factor, which equals zero for the samples of  cylindrical geometry, and $V$ is the entire sample volume as it was in the pure thermodynamic approach. 

Hence, taking into account that $r_i$ is the root mean square value,  
implying that the currents in Fig.\,15 overlap, we arrive at the distribution of the averaged microscopic field, i.e. $B$, as schematically shown in the graph in Fig.\,15. Thus, the MW model explains both the validity of the solely thermodynamic approach as well as the reality of the penetration layer. 

Finishing our discussion of H. London's argument or the second dilemma of the London theory, we see that its erroneousness, as in the case of the first dilemma,  stems from the assumption of the  screening current, which, in particular,  leads to the appearance of an apparently non-physical excluded volume in this theory.

\section*{ACKNOWLEDGMENTS}

\tg{It is my pleasant duty to thank Dr. A.-M. Valente-Feliciano, Dr. A. Suter, Dr. T. Prokscha and Prof. C. Van Haesendonck for their crucial contributions into the LE-$\mu$SR project temporarily interrupted by the pandemic. I am deeply grateful to Prof. Chris Van Haesendonck for }
his hospitality and many years of support, and Prof. Joseph O. Indekeu for driving me into studies of superconductivity.  I am pleased to thank Prof. Tony Leggett for an encouraging response on a short version of the manuscript. Special thanks to Prof. Vladimir Z. Kresin for important recommendations after reading an original version of the manuscript and for \tb{the invitation} to write this review.

\end{document}